\documentclass[12pt]{article}

\usepackage{times} 

\usepackage[authoryear,sort&compress]{natbib}

\defcitealias{IMF_tracker}{IMF, 2020}
\defcitealias{HACH_2020}{Hale et al., 2020}

\makeatletter
\def\@mb@citenamelist{cite,citep,citet,citealp,citealt,citeplatex,citetalias}
\makeatother

\usepackage{multibib}  
\newcites{latex}{Appendix References}

\usepackage{geometry}
\usepackage{afterpage}
\usepackage{graphicx}
\graphicspath{ {./images/} }
\usepackage{float}
\usepackage{amsmath}
\usepackage{enumitem}
\usepackage{caption}
\usepackage{subcaption}
\usepackage{url} 

\usepackage{amsthm,letltxmacro}
\usepackage{textcomp}
\usepackage{gensymb} 

\usepackage{cleveref} 
\crefname{theor}{hypothesis}{hypotheses}

\usepackage{changepage}
\LetLtxMacro\oldproof\proof
\let\endoldproof\endproof
\renewenvironment{proof}[1][\proofname]
  {\begin{adjustwidth}{2em}{2em}
   \oldproof[#1]}
  {\endoldproof
   \end{adjustwidth}}

\newtheorem{definition}{Definition} 
\newtheorem{theor}{Hypothesis} 
\newtheorem{findingA}{Result} 
\newtheorem{findingI}{Result} 

\usepackage{amssymb}
\usepackage{pifont}
\usepackage{xparse}
\usepackage{adjustbox}
\usepackage{longtable}
\usepackage{multirow}
\usepackage{booktabs} 
\usepackage[flushleft]{threeparttable} 
\usepackage{threeparttablex}
\usepackage{array}
\usepackage{rotating}

\usepackage[bottom]{footmisc}

\newcolumntype{P}[2]{%
  >{\begin{turn}{#1}\begin{minipage}{#2}\small\raggedright\hspace{0pt}}l%
  <{\end{minipage}\end{turn}}%
}
\usepackage{multicol}
\usepackage{xcolor,colortbl}

\definecolor{Gray}{gray}{0.85}
\definecolor{LightGray}{gray}{0.98}

\newcolumntype{a}{>{\columncolor{Gray}}c}
\newcolumntype{d}{>{\columncolor{LightGray}}c}

\newcommand\blfootnote[1]{%
  \begingroup
  \renewcommand\thefootnote{}\footnote{#1}%
  \addtocounter{footnote}{-1}%
  \endgroup
}

\usepackage{setspace}

\title{Social distancing in networks:\\A web-based interactive experiment}

\author{Edoardo Gallo$^{*}$, Darija Barak, Alastair Langtry}

\begin{document}
\maketitle

\blfootnote{\noindent 
Declaration of interest: none. Authors affiliation: University of Cambridge, Faculty of Economics, Sidgwick Avenue, Cambridge, UK. Edoardo Gallo and Darija Barak contributed equally to this work. Informed consent was obtained from all participants, and the research was approved by the Faculty of Economics Ethical Committee at the University of Cambridge (No: UCAM-FoE-20-01). The authors are grateful to Matthew Elliott, Freya Jephcott, Frederic Moisan, Gabriella Santangelo, Flavio Toxvaerd and two anonymous referees for helpful comments and suggestions, and to Ilia Shumailov for technical assistance. Thanks to seminar participants at the Large Scale Experiments Webinar and Cambridge-INET Networks Reading Group. Edoardo Gallo acknowledges support from the Cambridge Endowment for Research in Finance, and the Cambridge Humanities Research Grant Scheme. Darija Barak acknowledges support from the Economic and Social Research Council (ESRC) and the Royal Economics Society. Alastair Langtry acknowledges support from the ESRC. This work is the sole responsibility of the authors, and does not necessarily represent the official views of any of the funding agencies.
}
\blfootnote{$^{*}$ Corresponding author: edo@econ.ac.ac.uk (E. Gallo)}

\begin{abstract}
\doublespacing
    Governments have used social distancing to stem the spread of COVID-19, but lack evidence on the most effective policy to ensure compliance. We examine the effectiveness of fines and informational messages (nudges) in promoting social distancing in a web-based interactive experiment conducted during the first wave of the pandemic on a near-representative sample of the US population. Fines promote distancing, but nudges only have a marginal impact. Individuals do more social distancing when they are aware they are a superspreader. Using an instrumental variable approach, we argue progressives are more likely to practice distancing, and they are marginally more responsive to fines.
\end{abstract}



\clearpage
\doublespacing

In 2020, the COVID-19 pandemic caused one of the most significant global disruptions since World War II \citep{K_2020, ERT_2020, CMPRR_2020, Polyakova27934}, and paralyzed the global economy for over a year. To prevent healthcare systems from being overwhelmed and reduce the number of deaths \citep{M_2020, VKB_2020, Braunereabd9338}, governments worldwide enacted a range of Non-Pharmaceutical Interventions (NPIs). In particular, regulations to minimize social interactions by limiting exposure to others (social distancing) and/or self-isolating have emerged as a primary policy tool to manage infection levels in the short- and medium-term \citepalias{IMF_tracker, HACH_2020}.

Widespread adoption of, and compliance with, social distancing guidelines continues to be unprecedented. Throughout the pandemic, governments have used different policy tools to promote social distancing behavior. Some countries imposed heavy fines on anyone found breaching social distancing measures. For instance, during 2020, Singapore punished first-time offenders with a fine of up to 10,000 SGD (approx. 7,000 USD), and repeated violators could be jailed \citep{SMH_2020}. At the other end of the spectrum, India and the UK relied heavily on informational messages (nudges hereafter) early on, but then introduced fines after violations turned out to be common \citep{H_2020, S_2020}. 

Interestingly, the development and adoption of various COVID-19 vaccines has not put an end to the use of NPIs. The SARS‑CoV‑2 virus continues mutating, leading to the emergence of new variants of the disease which are resistant to the vaccines \citep{andrews2022covid,vogel2022new}. Moreover, some governments, such as China, continue to adhere to strict COVID-19 policies including mandatory lock-downs of entire cities \citep{forbes_22}. The reality is that social distancing policies continue to be used worldwide to combat COVID-19. However, even today governments lack robust scientific evidence on the relative effectiveness of different policies aimed at encouraging social distancing practices.

We conduct an interactive online experiment to investigate the effectiveness of fines and nudges as policy interventions in promoting social distancing, and the role of network position and the contagiousness of the disease. In the fine treatment, subjects pay a fine if they do not practice social distancing. In the nudge treatment, subjects watch a 3-minute video that explains the harm to others of not distancing. To investigate the role of network position, we assign participants to either a 5-node complete network (where everyone is connected to everyone else) or a 5-node star network (where participants are only connected to a central ``superspreader''). Finally, we vary the contagiousness of the disease by assigning participants to either a high or a low contagion treatment. The subject pool, recruited through Amazon Mechanical Turk (henceforth, MTurk), is a near-representative sample of the US population in terms of age, gender, and geographical location. All sessions took place in May 2020 at the height of the first wave of the COVID-19 pandemic. 

Our main finding is that fines significantly increase the level of social distancing by participants, while the impact of nudges is marginal. This supports the increased use of financial penalties by many governments, rather than just relying on informational messages as in the early stages of the pandemic \citep{M_2020, H_2020}. 

Second, network position matters. Superspreaders practice distancing more than both poorly connected peripheral individuals and individuals with the same number of interactions in a homogeneous group. Superspreaders played a pivotal role in the COVID-19 pandemic, but, to our knowledge, research has been largely limited to the biological dimension \citep{AWWLTCLC_2020, LGTBNL_2020}. There is, however, also a social dimension to being a superspreader -- individuals with many social interactions are more likely to spread the virus. Individuals are aware of being a superspreader in terms of social interactions, and our findings show that this leads to behavioral responses that are absent in the biological realm.

Third, political conservatives are less likely to practice distancing. The magnitude of the effect is significant and comparable to the impact of introducing a fine. We use an instrumental variable approach to provide evidence that this relationship is causal, and therefore the causal evidence here clearly relies on stronger assumptions than for the first two key results. This is consistent with the well-documented partisan divisions that characterize the COVID-19 response in the US \citep{ABCGTY_2020,gollwitzer2020}.

\textbf{Related literature.} This study has immediate methodological implications for policy-makers considering non-pharmaceutical interventions to contain the COVID-19 pandemic. It also contributes to both economics and epidemiology literatures. Below, we briefly summarize our contributions to these areas.

Our paper proposes a game that is novel in the experimental economics literature to investigate social distancing decisions. It is related to the standard public good game because it explores a trade-off between a costless self-interested decision and a costly one that benefits everyone in the group \citep{ledyard1994public}. The increase in social distancing in the fine intervention is analogous to the increase in contributions when there is a punishment mechanism in a public good game \citep{fehr2000cooperation}. Financial punishment for noncompliance with a policy has been studied in the behavioral and experimental economics literature in a variety of contexts. Importantly, a strand of this literature has documented that fines can backfire in some circumstances, primarily by crowding out intrinsic incentives \citep{gneezy2000fine, gneezy2000pay, gneezy2011and}. For this reason, it is important to collect evidence on the likely effects of such fines before their deployment. Our paper provides evidence on the effectiveness of fines to ensure compliance with social distancing in the context of COVID-19.

When it comes to policy interventions and their effect on individual behavior, a growing literature investigates the effectiveness of nudges. The attractiveness of nudges, especially in the form of informational messages, is that they are a relatively inexpensive tool that can be implemented and scaled easily. Yet, the evidence on the effectiveness of nudges from existing research is mixed. For example,a meta-analysis by \cite{hummel2019effective} finds that in empirical nudge studies only 62\% of nudging treatments are statistically significant. Hence, testing the effectiveness of nudges may be useful before rolling them out in the population. When it comes to the COVID-19 pandemic, there is some evidence that informational messages \emph{do} have an impact in certain settings. For example \cite{NBERw27496} test an informational message in West Bengal, India and find a significant impact on behaviour -- i.e. reporting of health symptoms to community health workers and travel beyond one's village. Our paper contributes evidence of the efficacy of such messages in the US.

A further contribution of our work is to show how awareness of being a superspreader affects behavioral responses. The role of superspreaders has received a lot of attention in the epidemiological literature, but studies are limited to the biological component where the individual is unaware of being a superspreader \citep{AWWLTCLC_2020, LGTBNL_2020} and/or empirical studies that correlate network position with health outcomes \citep{chen2021nursing}. In our work, instead, we show that position in the network \emph{causes} changes in social distancing behavior, and we can investigate how these behavioral responses vary with policy interventions.

In political economy, recent studies have investigated the impact of political ideology on social distancing decisions. For instance, both \cite{gollwitzer2020} and \cite{ABCGTY_2020} use sociomobility data from smartphones to measure physical distancing. Looking at county-level data, they find that Republicans do significantly less social distancing than Democrats. In contrast, we look at decision-making at the individual level, which allows us to relate individuals' political leanings to their behavior. Our findings are strongly complementary to the existing literature -- we also find that conservatives do significantly less distancing.  

Aside from the above key contributions, we also contribute to theoretical economics. A crucial new component of the social distancing game in our study is that decisions are affected by stochastic factors in the environment -- the contagion process and the partial effectiveness of social distancing. This relates our paper to experimental work on public goods with stochastic elements. In particular, \cite{FRD_2012} look at repeated play in a prisoner’s dilemma where intended actions are implemented with noise. Further, \cite{CFMJS_2014} investigate how the uncertainty about network structure affects subjects' ability to coordinate on efficient outcomes in games of strategic substitutes and complements. Unlike these works, in our social distancing game stochasticity originates from a probabilistic contagious process through which the disease spreads across the network. While using a contagious process is not entirely new to experiments on self-protection decisions (see, for example, \cite{chen2013behavioral,bohm2016selfish}), our study is, to the best of our knowledge, the first to incorporate an explicit network structure into the setup. As discussed above, this allows us to causally investigate the effects of the architecture of social connections in the group on individual behaviour and aggregate outcomes.

Field experiments have become a prominent methodology to test policies because they provide clean evidence of causality in real settings \citep{duflo2020field}. The COVID-19 pandemic has largely deprived policymakers of the ability to deploy field experiments due to movement restrictions including self-isolation, social distancing, and travel bans. In the last decade, several studies have run interactive web-based experiments in other disciplines \citep{rand2011dynamic, shirado2017locally} and, in a more limited fashion, within economics \citep{jackson2014culture, GY_2015}. Importantly, to the best of our knowledge, our work is the first to exploit the diversity of online subject pools to achieve representativeness along some sociodemographic dimensions (i.e. gender and geographic place of residence). Recent studies have shown that representative samples can differ from student populations with respect to fundamental behaviors (see, for example, \cite{chapman2022looming} on loss-aversion). Using a near-representative sample allows us to get a more informed perspective on the behaviors in the US population. Another crucial contribution of our work is to show how interactive web-based experiments are a novel methodology to test policies that complements existing ones and has distinct advantages. The final section of this paper contains further discussion on how interactive web-based experiments can complement and enhance studies based on sociomobility data and surveys.

The epidemiology literature about social distancing focuses on its macro-level impacts, i.e. how it might affect the evolution of the pandemic. Consequently, studies are mostly simulation-based. For example, \cite{fenichel2011adaptive} explicitly incorporate optimizing behavior by a representative agent into the classic SIR model and use simulations to examine how this affects disease dynamics. More recently, there has been a profusion of studies focused on the COVID-19 pandemic. \cite{K_2020} and \cite{chang2020modelling} both simulate disease dynamics for COVID-19, and explore the effects of a variety of non-pharmaceutical interventions on the predictions. In contrast, we elicit actual behavioral responses to counterfactual policies. Aside from their direct relevance, our findings can also be used to refine simulation-based epidemiological models by informing the choice of behavioral rules. 

The remainder of this paper is structured as follows. Section \ref{sec:exp_des} describes our experimental design. Section \ref{sec:data} summarizes our methods of data collection and presents the resulting sample. Section \ref{sec:results} presents the main results from our experiment, while Section \ref{sec:covariates} discusses covariates of social distancing decisions. Section \ref{sec:interpretation} interprets the results from the previous two sections. Finally, Section \ref{sec:remarks} contains a discussion of our results and the role of interactive online experiments in informing policy.

\section{Experimental Design}\label{sec:exp_des}

This section describes the game, the workflow of the experiment, and treatments.

\begin{figure}[t]
    \centering
    \includegraphics[width=\textwidth]{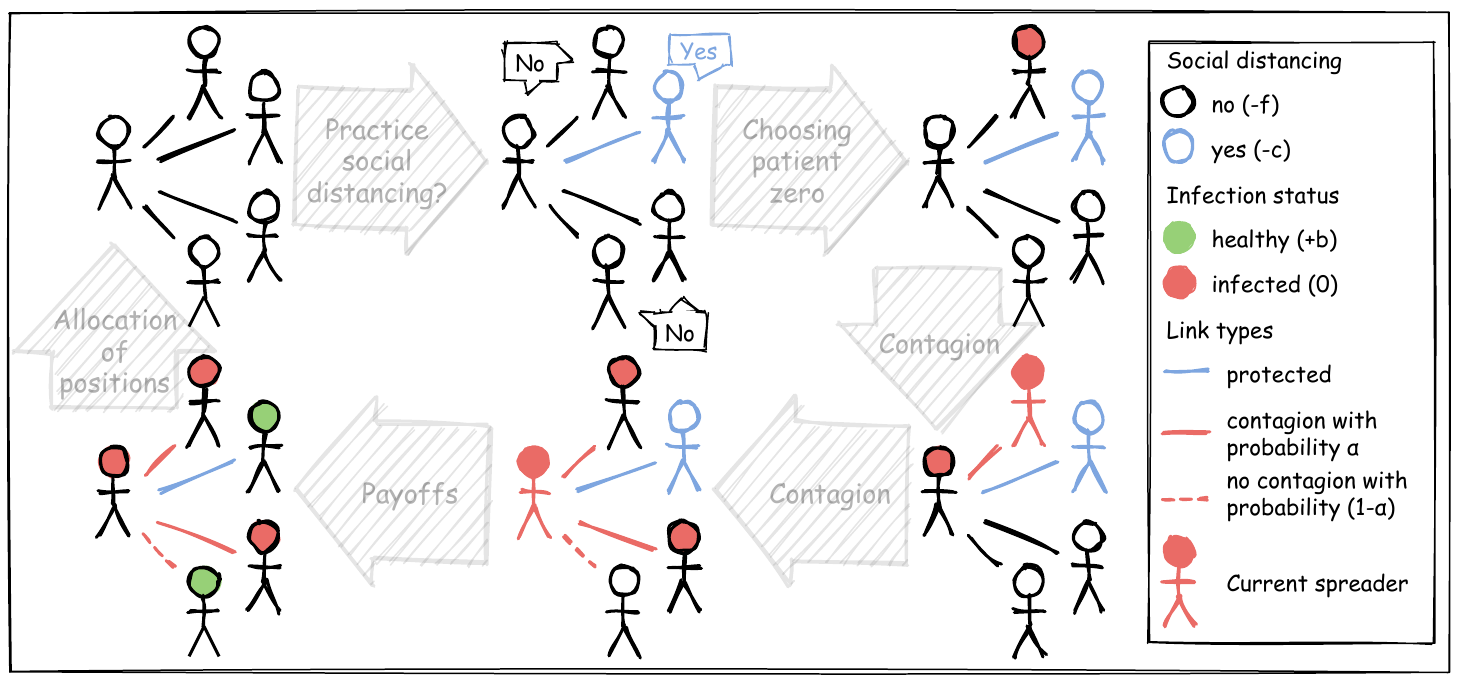}
    \caption{Flow of a typical round of baseline and intervention. In the experiment, we use the following parameterization: $f = 0$ or $15$ points (fine intervention only); $c = 35$ and $b = 100$ points; $\alpha = 0.15$ (low contagion) or $0.65$ (high contagion). Final payoffs for the round are a combination of individual social distancing choice and infection status. For example, a participant who practices social distancing and is healthy, receives $(-c+b) = 65$ points. In the figure, the chosen network architecture is the star. In the experiment, half of the treatments had the star network architecture and the other half had the complete.}
    \label{fig:round_flow_main}
\end{figure}

\textbf{Game.} \Cref{fig:round_flow_main} illustrates the social distancing game which corresponds to a round of the experiment. At the beginning of a round, participants are randomly allocated to one of the positions in an unweighted and undirected network (top left). The structure of the network is common knowledge. Subjects must simultaneously decide whether to practice social distancing at a known cost $c>0$ (top center). After the decisions are made, one and only one participant -- \emph{patient zero} -- is randomly picked to contract COVID-19 directly. In the example in \Cref{fig:round_flow_main}, patient zero is the participant highlighted in red in the top right panel. 

There are two benefits to practicing social distancing. First, if the individual is randomly picked to be patient zero, then she becomes infected with $50\%$ probability, rather than for sure. Second, any individual who practices distancing cannot transmit COVID-19 to others or contract it from infected individuals. In the example in \Cref{fig:round_flow_main}, patient zero decided to not practice social distancing, hence she becomes infected.

COVID-19 then spreads through the network from infected to healthy individuals who do not practice social distancing through contagion with a known probability $\alpha \in (0,1)$. The bottom panels of \Cref{fig:round_flow_main} show a possible spread in the example with three individuals infected and two healthy ones at the end of the contagion process.\footnote{The algorithm that we use is iterative and relies on the idea that each of the infected agents interacts with other agents exactly once. As a result, in a star network there may be at most two stages of contagion, whereas a complete network with five nodes permits at most four stages (for further details see Online \Cref{sec:model}).} At the end of the round, healthy individuals receive a benefit of $b=100$ points, whereas those infected receive zero benefit. Any individual who chose to practice social distancing pays the cost $c=35$ points, irrespective of their final infection status.

Assuming self-interested fully rational agents, the model predicts that the pure strategy Nash equilibrium outcomes will be inefficient for a wide range of parameter values. In fact, given an arbitrary network, it is possible to fully characterize both equilibrium and efficient outcomes.\footnote{See Online Appendix for the full characterization.} In particular, we show that, depending on the network structure, the equilibrium number of agents practicing social distancing can be below the efficient number. In this model, social distancing is essentially a merit good so it exhibits positive externalities -- welfare loss may therefore occur in the absence of intervention.

\textbf{Workflow.} In the experiment participants first play 20 rounds of the baseline game above in fixed groups of five subjects, with positions on the network being randomly reallocated in each round. Notice that the participants are not informed about the decisions and outcomes of other members of their group at any point during or after the experiment. After the initial 20 rounds, their group is treated with either the fine or the behavioral nudge policy intervention. In the fine treatment, whenever a participant decides not to practice social distancing she receives a fine of $f = 15$ points, independent of her infection status at the end of the round. In the nudge treatment, every participant must watch a 3-minute video explaining how failing to practice social distancing may harm others. The same group of 5 participants then play 20 rounds of the game under one of the policy interventions. Hence we investigate the impact of each policy on social distancing behavior using a within-subjects design and the relative effectiveness of the two policies with a between-subjects design. Once participants have completed the interactive part of the experiment, they proceed to the post-experimental questionnaire, with basic demographics questions and a set of knowledge and attitudes questions on a range of topics including the COVID-19 pandemic, social distancing, religion, global warming, ideology, and political affiliation. Finally, participants complete an incentivized Bomb Risk Elicitation Task (BRET) \citep{CF_2013} to elicit risk preferences. Subjects are explicitly primed to think about COVID-19 in the instructions by naming the disease and describing the main symptoms according to the guidelines by the Centers for Disease Control and Prevention.\footnote{Full instructions for the experiment are available at: https://github.com/darijahalatova/distancing\_instructions.}

\textbf{Treatments.} The focus of our experiment is on three treatment dimensions. First, we consider two 5-node network architectures for the structure of interactions among participants: (1) a complete network, where everyone is connected to everyone else, and (2) a star network, where one node is connected to all other nodes which are not connected between themselves. Next, we vary the level of contagiousness of COVID-19 to be either low ($\alpha = 15\%$) or high ($\alpha = 65\%$). Finally, we focus on two intervention methods: (1) a fine $f = 15$ points for not practicing social distancing, or (2) a behavioral nudge in the form of an informational video highlighting the harm caused to others by not practicing social distancing. Consequently, we have a $2 \times 2 \times 2$ full-factorial design, with a total of 8 treatments. 

We choose to study the effect of an informational video for two main reasons. First, as a (relatively) cheap, easy to deploy, and non-invasive intervention, it is a natural benchmark against which to compare other policies. To justify more expensive and onerous measures -- such as fines --  policymakers ought to want those measures to be more effective than a nudge, not just more effective than doing nothing. Second, governments across the globe used informational messages extensively during the height of the COVID-19 pandemic.\footnote{For example, the UK Government spent over £240 million on COVID-19 advertising in 2020 alone \citep{MBA_21}.} Understanding whether these interventions are likely to work is therefore important in its own right. Additionally, while a pure informational message has no impact in the standard rational agent framework (which our model assumes for simplicity), a range of studies find that they can affect behavior \citep{NBERw27496, barak2022experience}. 

The complete and star networks we use as treatments are stark, but they are a useful starting point for investigating the role of the network. They present clearly differentiated structures, making them useful for identifying causal effects.\footnote{They are also common choices for networks experiments (see, for example, \cite{frey2012equilibrium,rosenkranz2012network, choi2014communication}). However, if responsibility for protecting others is more diffused in less stark networks, it is possible that network structure plays a lesser role there. \cite{falk2020diffusion} presents evidence suggesting that people are more willing to take actions seen as immoral if their role in decisions is not pivotal.} Additionally, they allow tractable analytic solutions to our model and hence sharper theoretical predictions (see Online \Cref{sec:model}).

\section{Data} \label{sec:data}

In this section, we summarize our data collection methods and present the resulting dataset. The section also includes a description of the key demographics of the final sample and a convergence analysis of the decision data from the experiment.

Following standard practices in interactive online experiments \citep{GY_2015, SW_2011}, we first recruit a standing panel of subjects using a short survey on MTurk. During recruitment, we use the 2018 US Census data to generate a representative panel of the adult US population in terms of age, gender, and geographical location \citep{census_2019}. As part of recruitment, we also collect subjects' social preferences using the Social Value Orientation (SVO) task \citep{MAH_2011}. The SVO task classifies individuals into four categories -- prosocial, individualistic, competitive, and altruistic. This task was incentivized.

For each of the experimental sessions, we invite a random representative sample of subjects from our standing panel.\footnote{We restrict participation to US-resident participants who have completed at least 500 tasks and have an overall ranking of at least 96\%. We also ensure that no two subjects with the same IP address participate in the experiment. Finally, in line with recent contributions \citep{rand2011dynamic,gallo2015effects}, multiple participation is not allowed even by participants who have only seen the instructions.} We obtain a near-representative sample of $n=400$ participants completing the experiment. To check that our sample is balanced when it comes to treatment assignment, we run a chi-squared test on participants’ assignment to treatment (a categorical variable with 8 options) with respect to gender, age category, and geographical location variables. Results indicate that subjects in all eight experimental treatments are not significantly different from each other (two-sided $\chi^2$, $p >0.1$, see \Cref{tab:treatment_balance} for details).

The experiment took place between May 4th and 29th, 2020.\footnote{We used Qualtrics (www.qualtrics.com) for recruitment, and the main experiment was programmed in o-Tree (www.otree.org) \citeplatex{CSW_2016} with a server deployed on Heroku (www.heroku.com).} Each session had between one and five independent groups of five subjects playing simultaneously, with the exact numbers depending on turnout. The experiment lasted on average 33 minutes, and subjects earned an average of 5.81 USD (including a fixed participation fee of 1 USD). Subjects remained completely anonymous throughout the experiment, and informed consent was obtained from subjects before participation.

Prior to the experiment, we screened Mturkers in two ways. First, we only recruited Mturkers with 500 completed Human Intelligence Tasks (HITs) on MTurk \emph{and} an approval rate of at least 96\% into our standing panel. Second, we tracked IP addresses, and did not permit the same IP address to take part more than once.\footnote{As well as preventing a given person participating more than once, this also seeks to stop more than one person from the same household from participating. For these purposes, even viewing the experimental instructions counts as taking part.} Within the experiment, we screened participants in two main ways. First, we used a 3-question quiz to check their understanding of the instructions. Participants who failed three times were excluded before the experiment started. Second, we gave subjects 20 seconds to make each social distancing decision, and we disqualified them if they missed three consecutive decisions.  Throughout the experiment, 6 subjects were disqualified for this reason. We dropped all data from their respective groups from our analysis -- but adding it back in has no material impact on any of our results. \ref{sec:data_quality} of the Online Appendix provides further detail on these, and other steps we took to ensure data quality. 

All 400 subjects in our sample are residents of the US, 47\% are female, and the mean age is just under 44 years. The average subject in our sample has 15.3 years of education and 70\% are employed (either full or part-time). In terms of religious attitudes, 51\% of our participants consider themselves religious, with 39\% reporting Christianity as their primary religion. The average subject in the sample is moderately risk-averse, with a CRRA score of $0.53-0.55$. With respect to social values, our sample mostly consists of prosocials (58\%) and individualists (41.8\%), and only one subject is classified as competitive.\footnote{See Online Appendix for a more detailed description of the demographics of the sample.} 

Overall, the final dataset contains 40 decisions for each of the 400 subjects for a total of 16,000 observations. As subjects played simultaneously in groups of five, there are 80 distinct clusters (i.e. groups). \Cref{fig:1}A shows the evolution of average individual propensity to social distance aggregated over contagiousness and network treatments. The figure shows a downward trend in social distancing behavior in the first rounds. Such behavior is typical for public goods experiments whereby contributions to the public good decline as the session progresses \citep{ledyard1994public}. In this paper, we are interested in the limiting outcomes of this convergence process, and therefore we would like to restrict our attention to those rounds where the majority of subjects have settled on some stable strategy.

In particular, we define that a participant has converged to a stable strategy by a certain round if she used this strategy for the previous four rounds, and in all subsequent rounds, the number of consecutive deviations from the chosen strategy does not exceed two.\footnote{We consider three types of convergence strategies. In both networks, we look at the strategy where the subject always chooses the same action. For the star network, we also consider two extra strategies. In one strategy the participant always chooses the same action when she is in the superspreader position and the complement action when she is peripheral. In the other strategy, she always chooses the same action when she is the superspreader and alternates between the actions when peripheral.} Given this convergence definition, we find that at least $80\%$ of participants converge to a stable strategy in every treatment after the first 10 rounds. Our main analysis, therefore, focuses on the last 10 rounds of the baseline and intervention in order to control for the time trend due to learning and experimentation by subjects. The results are robust to including all of the data. 

\section{Determinants of social distancing decisions}\label{sec:results}

This section investigates the determinants of social distancing decisions, using both non-parametric and parametric methods. The analysis in this section is based entirely on our treatment variables -- policy intervention (fine or nudge), network architecture and the rate of contagiousness of the disease. 

The non-parametric analysis uses the Mann-Whitney U-test (MW hereafter) \citep{MW_1947} for unmatched samples and the Wilcoxon Signed-Rank test (WSR hereafter) \citep{W_1992} for matched samples. Samples are aggregated at the group level and matched when they are generated by the same groups, and they are  unmatched when making comparisons across groups from different treatments.\footnote{All results using these non-parametric tests are robust to using their parametric analogs, i.e. t-test for unpaired and paired samples (see \cref{sec:analysis} of the Online Appendix).}

Our parametric approach relies on the Random Effects Logit (henceforth, REL) to model the binary choice of whether to practice social distancing. Formally, it models a random utility model with individual-specific random effects: 
\begin{align}
   y_{it} = 1 \iff x_{it} \beta + v_{i} + \epsilon_{it} > 0 
\end{align}
where $y_{it}$ is the decision to practice social distancing, $x_{it}$ is the set of controls, $v_{i}$ are individual specific random-effect and $\epsilon_{it}$ is an error term with a logistic distribution. It allows individual-specific propensity to practice social distancing, but assumes that the impact on utility of a variable $x$ is not individual-specific. We cluster errors at the group-level. Our main specification (column 5 in \Cref{tab:main text table}) includes the following five categories of independent variables $x_{it}$: (1) dummy variables for the experimental treatments (fine intervention, nudge intervention, contagion level, and node types), (2) demographics controls, (3) location and institutional controls, (4) preference controls, and (5) ideology controls. P-values throughout this section are presented for a single specification, but, as \Cref{tab:main text table} shows, results are robust across all specifications.\footnote{Further, the results are robust to using a Linear Probability Model or a Probit model. We discuss this further in the Online Appendix.}

A preliminary question is whether the policy intervention has an impact on the aggregate level of social distancing. We conduct a Wald test to identify a structural break. This test assumes that there is at most one structural break in the data, but is agnostic as to when it occurs and whether it occurs at all. In the fine intervention, a change in the evolution of social distancing occurs at round 21, immediately after the implementation of the fine (Wald test, $p<0.001$ for all data and subdivided by the other treatments). The same break at round 21 occurs in the nudge intervention in the aggregate data (Wald, $p<0.001$), but the result is not robust once we subdivide by the other treatments. See \Cref{tab:structural break} in the Online Appendix for a full breakdown. The introduction of the fine therefore has an immediate impact on social distancing behavior, while the impact of the nudge is less clear-cut. 

Further, we find that fines increase the level of social distancing, and the effect is significant both in the non-parametric (WSR, $p=0.0008$) and in the regression (Specification 1, S1 hereafter, $p=0.001$) analyses. The increase in social distancing behavior is present both in low (WSR, $p=0.04$) and high (WSR, $p=0.003$) contagion environments. In the last 10 rounds of the fine intervention, there is an $8\%$ increase in social distancing compared to the last 10 rounds of the pooled baseline so the effect of the fine is sizeable and long-lasting.

In contrast, the impact of the nudge intervention is marginal and not robust to all specifications (WSR, $p=0.008$; S1, $p=0.03$). Interestingly, narrowing the focus to a specific contagion environment, the nudge significantly increases distancing with high contagion (WSR, $p=0.0003$) but it has no effect with low contagion (WSR, $p=0.5$). Comparing the two policies, the impact of the fine is marginally higher than the nudge (MW, $p=0.08$). Overall, the increase in social distancing in the last 10 rounds due to the nudge intervention is only $2\%$.

A crucial driver in the spread of COVID-19 is the presence of superspreaders -- individuals who go on to infect a much larger number of others than the average \citep{AWWLTCLC_2020, LGTBNL_2020}. A primary determinant of being a superspreader is biological -- something that cannot be varied experimentally and is outside of the scope of this study. However, there is also a social element driven by the wide variations in the number of social interactions across individuals \citep{jackson2007meeting}. An important difference between biological and social drivers of being a superspreader is that an individual is aware of the latter, and therefore the tendency to social distance may vary with position in the network.

\begin{figure}[t]
\centering
\includegraphics[width=0.9\linewidth,keepaspectratio]{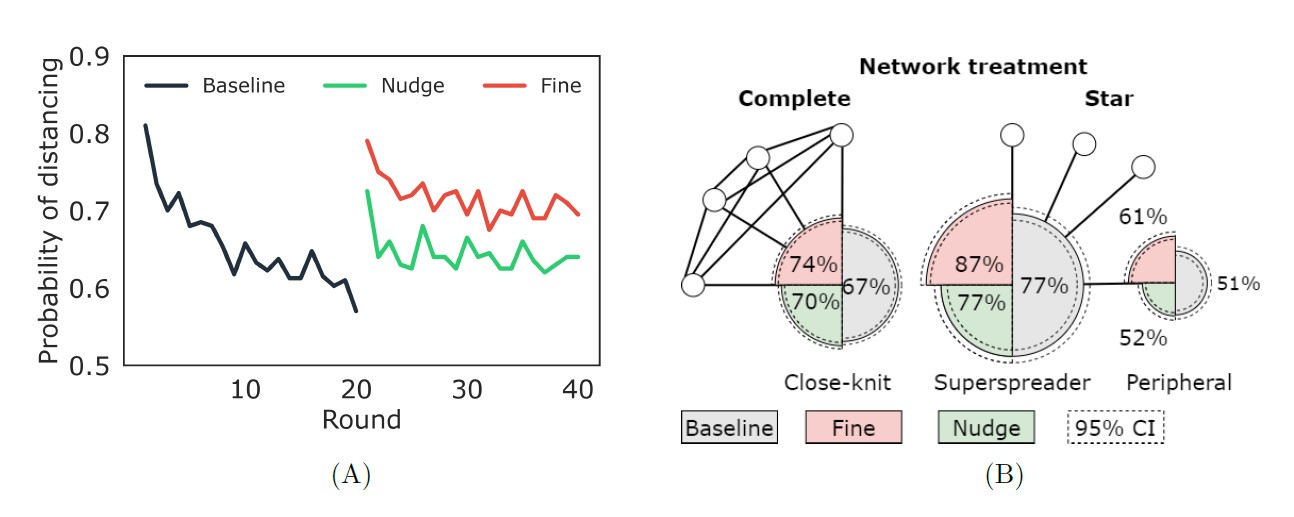}
\caption{Policy and network position. (A) Evolution of aggregate probability of distancing split by policy intervention. (B) Probability of distancing split by network position and policy intervention. Node size is proportional to distancing probability. Width of dotted band around the circumference of each circle indicates 95$\%$ confidence intervals.}
\label{fig:1}
\end{figure}

\Cref{fig:1}B illustrates the three positions in our experiment drawn with the node size proportional to the amount of social distancing in that position for each policy treatment. On the left side, there is the complete network in which all participants are in what we dub a `close-knit' position. On the right side, there is the star network with one participant -- the superspreader -- interacting with all others in what we dub a `peripheral' position. Superspreaders practice more social distancing than peripheral participants both in the baseline and policy interventions (MW and S1, $p<0.0001$ for all). Moreover, superspreaders also practice social distancing more than close-knit participants (MW, $p=0.004$; S1, $p=0.03$) despite having the same number of interactions. One potential driver of this behavior is that they are aware of their central role and want to protect the group. Another potential driver is that they realize peripheral participants are less likely to distance due to their isolated position. Our experimental design does not allow us to investigate the relative importance of these two drivers, but this is an interesting avenue for future research. High contagiousness leads to significantly more social distancing (MW, $p=0.01$, S1, $p < 0.0001$), and the size of the effect is similar to the difference between being in the close-knit and peripheral positions. Social distancing is $10\%$ and $15\%$ higher in the high contagion setting than in the low contagion setting in baseline and intervention respectively.   

\section{Covariates of social distancing decisions} \label{sec:covariates}
This section describes the association between social distancing decisions and a number of individual-specific variables including political ideology, risk attitudes, social preferences, and a range of socio-demographic characteristics. Section \ref{sec:ideology} focuses solely on political ideology where we use an Instrumental Variables (IV) approach to argue that the relationship is causal. It also discusses the strong identifying assumptions we make in the analysis. Section \ref{sec:other} describes other correlates of social distancing decisions, where we make no claim of causal effects. 

\subsection{Political ideology} \label{sec:ideology}
We find that self-reported Democrats are significantly more likely to practice distancing than Republicans (S4, $p=0.0002$). However, this binary classification is a very coarse measure. To obtain a more fine-grained picture of participants' political leanings, we construct an ideology index, with a score for each subject based on responses to questions about (1) support for President Trump’s handling of the COVID-19 pandemic, (2) support for universal healthcare, and (3) belief that social distancing measures impose unjustified economic costs. Questions are on a 5-point Likert scale, so we obtain a 0-12 index with 0 and 12 indicating extremely progressive and extremely conservative participants respectively.

\begin{figure}[t]
\includegraphics[width=\linewidth, keepaspectratio]{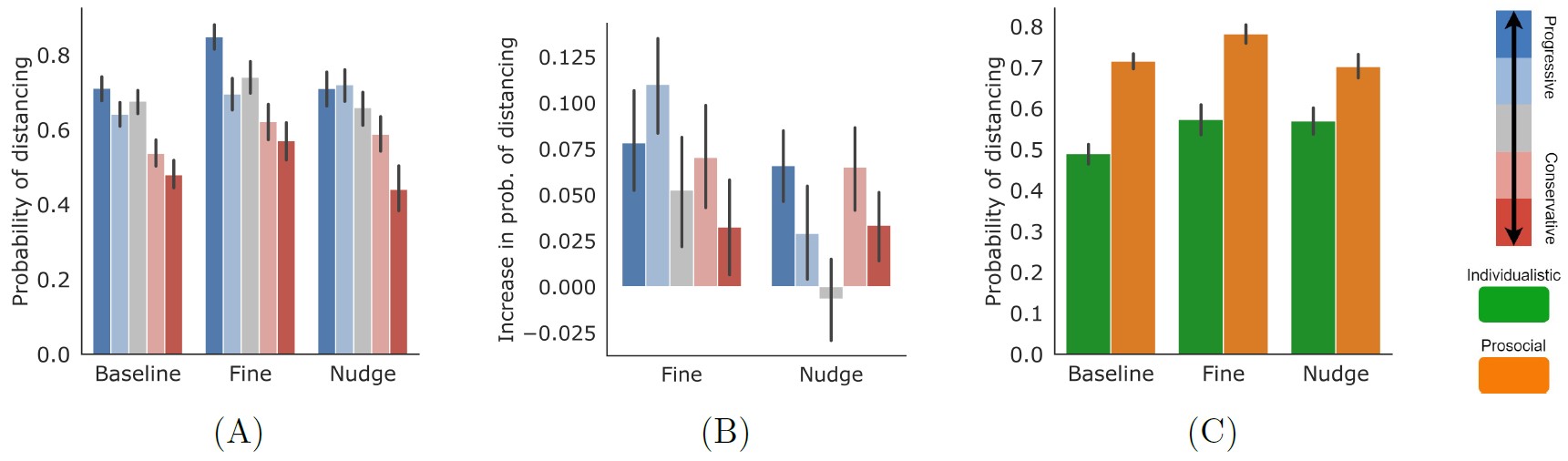}
\caption{Ideology and social preferences. (A) Probability of distancing and (B) Percentage change in probability of distancing split by policy intervention and quintile of ideology index. (C) Probability of distancing split by social preference and policy intervention.  Vertical bars indicate 95\% confidence intervals.}
\label{fig:2}
\end{figure}

\Cref{fig:2}A shows that the probability of practicing distancing decreases the more conservative a participant is -- an increase in ideology index from 1 to 5 (25th to 75th percentile among our subjects) corresponds to a $14.6\%$ decrease in the probability of distancing. The ideology index is a highly significant correlate of social distancing decisions (S5, $p<0.0001$), in line with recent sociomobility and survey-based studies in the US \citep{ABCGTY_2020, BH_2020, SSDB_2020}.

Our experimental design exogenously varies policy intervention, contagiousness, and the network, allowing us to make causal inferences on these dimensions. The same is obviously not the case with ideology.\footnote{In general, however, it is possible to incorporate subjects' ideology and other aspects of identity into experimental design. We thank an anonymous referee for pointing us towards examples of such work, where some features of subjects' identity are made salient \citep{benjamin2010social,chen2009group}.} We investigate the causal effect of ideology with an instrumental variable approach, using a measure of participants' skepticism of global warming as the instrument. As a partisan issue in the US, attitudes toward global warming are strongly correlated with political ideology \citep{HB_2016, PEW_2019}. The exclusion restriction our IV identification relies on is that global warming attitudes do not affect social distancing decisions separately from their association with ideology.

However, it is likely that global warming attitudes at least partly reflect geography and local institutions -- both of which might affect social distancing decisions. Failure to control for these would lead to violation of the exclusion restriction. We therefore add a set of proxy variables: federal regions indicator (see Online \Cref{sec:design} for details), population density, total number of COVID-19 cases to date, daily number of COVID-19 deaths, and an indicator of whether a stay-at-home order was in place.\footnote{Population density is at the county level. COVID-19 variables are at the state level. Cumulative number of COVID-19 cases, new daily COVID-19 deaths, and stay-at-home order indicator are for the date immediately before the day subject participated in the experiment. We do not report coefficients for them in Table 1 (we do so in \Cref{tab:logit_main} in the Online Appendix).} While these are necessarily imperfect proxies, it suggests that our instrument is not just capturing geography and local institutions.\footnote{Following a comment from an anonymous referee, we further checked that ideological composition of the state of residence -- as captured by the political party affiliation of the governor as of May 2020 -- and the level of religiosity -- measured by the share of people who report at least weekly worship attendance and those who pray daily \citep{pew2016} -- play no effect in determining distancing behavior. All results reported in this paper are robust to these checks.}

Further, it may be possible that our skepticism of global warming index is picking up subjects' general attitudes towards science and trust in information from experts. In turn, lack of trust in science may have an effect on subjects' social distancing behavior. While such concerns cannot be completely eliminated, existing research suggests that whereas political conservatism is strongly associated with skepticism towards global warming, understanding of science (or lack thereof) is not a climate change acceptance predictor \citep{rutjens2018not}.

Specification 6 (S6) in \Cref{tab:main text table} is identical to our main specification (S5), except that it uses an instrument for ideology. The instrument is an index of climate change attitudes -- constructed from subjects' beliefs that global warming is (1) happening, (2) caused mostly by humans, and (3) affecting weather in the USA.\footnote{We used the questions from \cite{HMML_2015}.} Political ideology is a significant causal determinant of social distancing, with conservatives less likely to practice distancing (S6, $p=0.04$). \Cref{fig:2}B illustrates an interesting further interaction between political ideology and the type of policy intervention. While the nudge intervention has no differential impact on subjects with different ideologies, conservative participants are marginally less responsive to the fine (S6, $p=0.08$). Note that using the ideology index itself (as in specification 5) instead of the IV approach (as in specification 6) has very little effect on the results.

\subsection{Other covariates} \label{sec:other}
Social preferences play an important role in the decision to social distance. First, an analysis of equilibrium decisions with self-interested individuals predicts levels of social distancing that are significantly lower than participants' decisions in all treatments (MW, $p<0.0001$). Second, using the SVO task, we classify subjects into individualistic and prosocial. As \Cref{fig:2}C illustrates, prosocial participants are $20\%$ more likely to choose distancing (S2, $p<0.0001$).  

Next, we elicit participants' risk preferences using BRET. As expected, risk-seeking individuals are less likely to practice social distancing (S2, $p=0.001$), and the effect is about half the size of being prosocial in the SVO task. The post-experimental survey includes a question asking why subjects choose to stay at home -- participants who list ``protecting others'' as one of the stated reasons are more likely to social distance (S2, $p = 0.007$), and the effect is similar in size to being risk-averse.

In terms of sociodemographics, females (S2, $p=0.004$) and older individuals (S2, $p<0.0001$) are more likely to distance. From the data, being female translates to a $12.6\%$ increase in the probability of distancing, and there are no significant interactions between gender and the effect of each policy intervention. Social distancing is increasing with age -- an average 60-year-old is $40.1\%$ more likely to distance than an otherwise identical 20-year-old. Interestingly, older subjects are relatively more responsive to the nudge (S3, $p=0.02$) and relatively less responsive to the fine (S3, $p < 0.0001$). Whites are less likely to distance (S2, $p=0.04$), but the effect is not robust. Education, religion, and employment status are not associated with distancing behavior. These associations between sociodemographic characteristics and distancing are broadly consistent with evidence from sociomobility \citep{BMB_2020} and survey-based data \citep{MSBAKMHLW_2020, GPPBBF_2020, PF_2020, PZB_2020}.

\section{Interpreting the results} \label{sec:interpretation}
The nonlinearity of the Random Effects Logit model makes interpreting the point estimates in \Cref{tab:main text table} difficult. The partial effect of any given variable depends on both the initial value for that variable and on the values of all variables. For a given individual, the partial effect of changing a variable from $x$ to $x'$ is not the same as that of changing it from $x'$ to $x''$. As a consequence, the estimated partial effect is different for every person.

To meaningfully compare the effect of different factors, we compute the Average Partial Effects (APE) calibrated on the characteristics and/or decisions of the subjects in our experiment. To do this, we calculate the partial effect for each participant in the experiment, and then take the simple mean over these individual-specific partial effects. For the individual-specific partial effect, we compare the estimated probability of distancing for each subject for two different values of the variable of interest while keeping all other variables at their observed value for that individual. For binary variables, we compare $0$ and $1$. For continuous variables, we compare the $25$th and $75$th percentiles (based on the experimental data). 

\begin{figure}[t]
    \includegraphics[width=0.9\linewidth, keepaspectratio]{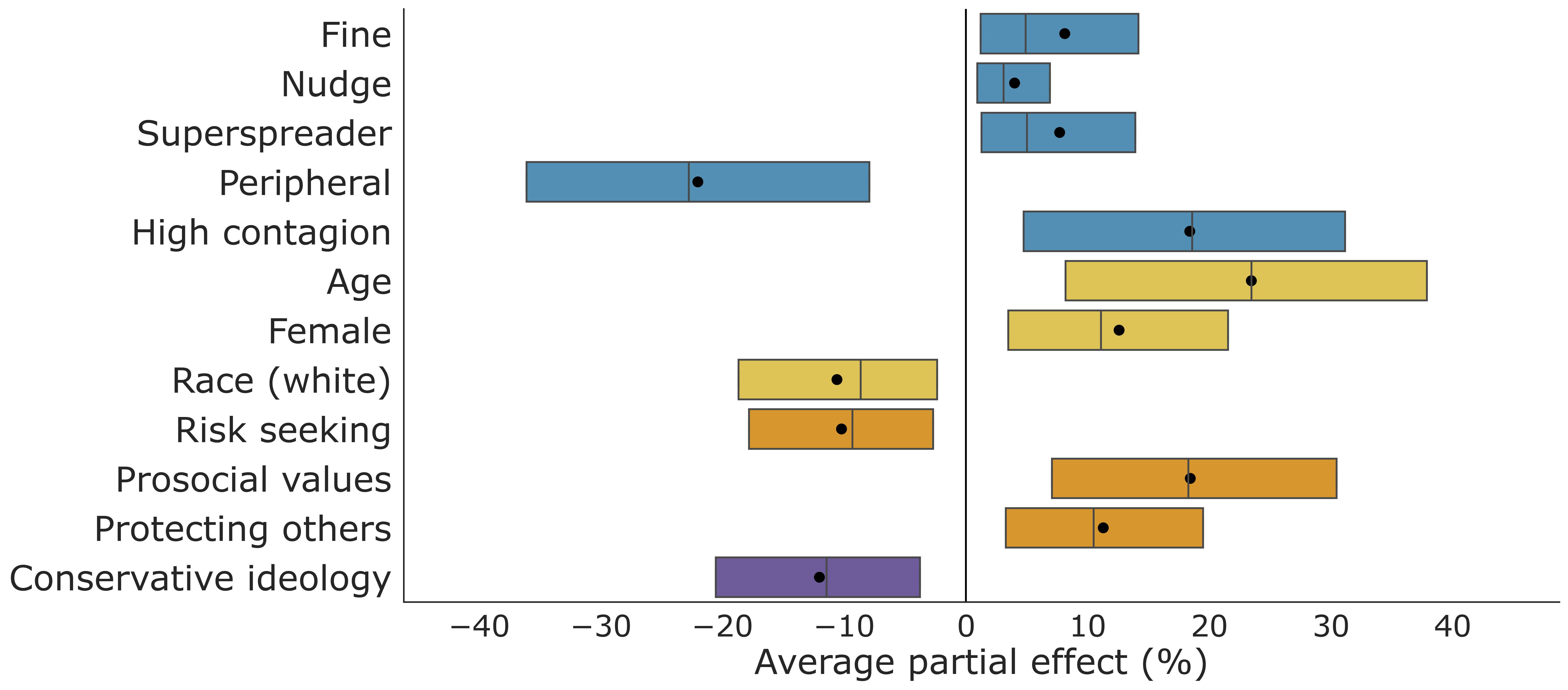}
    \caption{Average Partial Effects (APE) for variables that are significant in the REL main specification. Bar boundaries indicate 25th and 75th percentiles, vertical line inside bars indicate 50th percentile, and black dots show the mean. Blue bars indicate treatment variables. Orange and yellow bars indicate effects from social preferences and demographics respectively. Purple bar is the ideology variable. Variables not included because they are not significant include: education, religion, labor force participation, case numbers, and geography controls.}
    \label{fig:3}
\end{figure}

\Cref{fig:3} shows the mean partial effects (APEs) for all variables that are statistically significant in our main specification, as well as their 25th, 50th, and 75th percentiles. Blue bars indicate the impact of determinants of social distancing decisions (i.e. treatment variables). A high contagion setting and being in the peripheral network position have the largest impact with $18\%$ and $-22\%$ respectively. The fine intervention and being in the superspreader position have a moderate impact of about $8\%$. The smallest impact is the nudge intervention with only a $4\%$ effect.

The rest of the bars in \Cref{fig:3} refer to the covariates of social distancing decisions. Yellow bars indicate sociodemographics correlates. Increasing age from the 25th to the 75th percentile in our data has a $23\%$ effect on distancing. Gender and race have a moderate effect of around $10\%$. Orange bars display the effect of correlates related to preferences. Being prosocial is associated with a large $20\%$ effect on distancing, while risk preferences and the desire to protect others have a moderate impact of about $10\%$. Finally, the purple bar shows that political leanings have a moderate $12\%$ effect, which we argue is causal in our analysis.

Omitted from \Cref{fig:3} are a set of demographic, institutional, and geographic factors that are not statistically significant. A subject's level of education, religion, and labor force participation are all unrelated to their social distancing decisions. Also insignificant are their geographic location (as measured by Federal Region), the number of COVID-19 cases and deaths in their state, and population density. Finally, \Cref{fig:3} clearly shows how the partial effect of any given variable is highly heterogeneous across subjects. In fact, the range in partial effects from the 25th to 75th percentiles \emph{exceeds} the mean in many cases.

\section{Concluding remarks} \label{sec:remarks}

The sudden disruption brought about by the COVID-19 pandemic demanded a quick response from policymakers who had to implement novel social distancing policies whose success determined the fate of tens of thousands of lives. Our work provides some of the first experimental evidence on the efficacy of these policies and how it varies across social and individual characteristics. Additionally, it shows how the novel methodology of interactive web-based experiments can play a crucial role to inform policymakers, and complement data obtained using sociomobility and survey-based studies.

Our main finding is that fines are effective at increasing social distancing, while the impact of nudges is marginal.\footnote{This result shows an important benefit of fines. Governments should also take into account the costs of using fines when setting policy. These costs could include non-economic factors, such as \emph{per se} objections to intrusion on individuals' lives.} A limitation of our design is that we examine one specific form of nudge -- an information video that emphasizes the harms to others of failing to practice social distancing. However, the efficacy of nudges is sensitive to the content, media, and cultural contexts. For instance, \cite{NBERw27496} show that nudges in the form of 2.5-minute information videos are effective in encouraging social distancing in the Indian context. Despite their low impact in our study, future research should continue exploring the efficacy of nudges in promoting social distancing because their ease of deployment makes them a particularly attractive tool for policymakers. 

Another important funding is that conservative-leaning individuals are less likely to practice social distancing in the US. While this tendency has been documented in other empirical studies \citep{gollwitzer2020, ABCGTY_2020}, we provide evidence of a causal relationship using an IV approach. An open question is whether this relationship is peculiar to the US context or it extends to other countries with a less stark partisan divide. 

In terms of methodology, this study highlights the important role interactive web-based experiments can play in providing timely information to policymakers to test the effectiveness of policies. Experiments are the gold standard to test the causal impact of policy interventions, but lab and/or field experiments were severely constrained during the COVID-19 pandemic, depriving researchers of the standard tools to identify causal effects \citep{HM_2020}. Web-based experiments, however, were unaffected.
Moreover, they can be deployed quickly, scaled up at minimal cost, and easily reach a diverse sample and/or specific populations of particular interest. 

Evidence regarding behavioral responses to social distancing policies has so far been mostly based on either sociomobility data or survey studies based on self-reported claims about hypothetical behavior. Aside from a clean identification of causality, web-based interactive experiments complement these methodologies and present some further distinct advantages.

Mobility data provides detailed information on behavior in real-life settings that can verify the external validity of web-based experiments. For instance, recent studies validate our finding that conservative-leaning individuals practice less social distancing \citep{ABCGTY_2020, BH_2020}. However, it is usually stripped of any personal information, including sociodemographics, so that inferences about the effects of these variables can only be made for geographic units rather than individuals. Moreover, sociomobility data is unable to shed light on the effectiveness of counterfactual policies, while policymakers frequently need information in advance of implementation. 

Surveys are a standard methodology for fast, large-scale data collection, but our study raises some question marks about their accuracy in gauging actual behavioral responses. In our post-experimental survey, we ask participants to estimate how effective fines and nudges are in promoting social distancing. Participants significantly overestimate the efficacy of nudges compared to their actual choices in the experiment independently on whether they are in the fine (MW, $p<0.0001$) or nudge (WSR, $p=0.001$) treatment.\footnote{For further details of this analysis see Section \ref{sec:analysis_perception} of the Online Appendix.} This difference between survey-based self-reported intentions and actual behavior is consistent with recent evidence on the short-lived efficacy of nudges \citep{BFL_2017}. In other words, picking a policy based on self-reported measures may lead policymakers to opt for the wrong solution. These results caution against excessive reliance on survey-based self-reported data to gauge behavioral responses. 

\newgeometry{top=1cm,bottom=0.5cm}

\thispagestyle{empty}
\afterpage{\setlength\LTleft{-17.5mm}
\begin{table}[H]
\caption{Logit regression analysis of the experimental data.}\label{tab:main text table}
\hskip-2.0cm
\begin{tabular}{llcccccc}
\toprule
&  & \multicolumn{6}{c}{Decision to practice distancing (binary)} \\ \cline{3-8} 
& Variables                      & (S1)          & (S2)           & (S3)        & (S4)       & (S5) Main & (S6) IV        \\
\midrule
\multirow{10}{*}{\rotatebox{90}{\textbf{Treatment}}}  
    & Fine                       & 0.755***     & 0.761***      & 3.023***   & 0.942***  & 1.146*** & 1.147***   \\
    &                            & (0.231)      & (0.242)       & (0.515)    & (0.287)   & (0.352)  & (0.352)   \\
    & Nudge                      & 0.347**      & 0.343**       & -0.928     & 0.353     & 0.438*   & 0.437*    \\
    &                            & (0.157)      & (0.155)       & (0.582)    & (0.202)   & (0.261)  & (0.261)   \\
    & High contagion             & 1.330***     & 1.537***      & 1.561***   & 1.238***  & 1.547*** & 1.543***   \\
    &                            & (0.336)      & (0.323)       & (0.326)    & (0.387)   & (0.323)  & (0.322)   \\
    & Superspreader               & 0.975**      & 0.830**       & 0.845**    & 0.870*    & 0.833**  & 0.846**   \\
    &                            & (0.438)      & (0.421)       & (0.425)    & (0.484)   & (0.424)  & (0.429)   \\
    & Peripheral                 & -1.608***    & -1.780***     & -1.819***  & -1.760*** & -1.785*** & -1.771***  \\
    &                            & (0.322)      & (0.333)       & (0.338)    & (0.424)   & (0.333)   & (0.340)   \\ \hline
\multirow{6}{*}{\rotatebox{90}{\textbf{Preferences}}} 
    & Bomb risk score            &              & -0.0296***    & -0.0297*** & -0.0226** & -0.0296*** & -0.0300*** \\
    &                            &              & (0.00889)     & (0.00894)  & (0.0107)  & (0.00896) & (0.00885) \\
    & Prosocial values           &              & 1.485***      & 1.478***   & 1.570***  & 1.495***  & 1.503***   \\
    &                            &              & (0.334)       & (0.337)    & (0.365)   & (0.335)   & (0.334)   \\
    & Protect others             &              & 0.918***      & 0.984***   & 0.975**   & 0.926***  & 1.017**   \\
    &                            &              & (0.342)       & (0.342)    & (0.381)   & (0.346)   & (0.395)   \\ \hline
\multirow{6}{*}{\rotatebox{90}{\textbf{Demographic}}} 
    & Age (years)                &              & 0.0793***     & 0.0867***  & 0.0873*** & 0.0796*** & 0.0796***  \\
    &                            &              & (0.0137)      & (0.0142)   & (0.0164)  & (0.0139)  & (0.0139)  \\
    & Female = 1                 &              & 1.069***      & 1.070***   & 1.122***  & 1.072***  & 1.054***  \\
    &                            &              & (0.372)       & (0.370)    & (0.428)   & (0.372)   & (0.380)   \\
    & Race = White               &              & -0.960**      & -1.079**   & -0.941    & -0.955**  & -0.986**  \\
    &                            &              & (0.469)       & (0.468)    & (0.680)   & (0.468)   & (0.464)   \\ \hline
\multirow{4}{*}{\rotatebox{90}{\textbf{Ideology}}}    
    & Conservative index         &              & -0.282***     & -0.278***  &           & -0.250*** & -0.207**   \\
    &                            &              & (0.0565)      & (0.0567)   &           & (0.0591)  & (0.0995)  \\
    & Republican dummy           &              &               &            & -1.529***  &           &            \\
    &                            &              &               &            & (0.403)   &           &           \\ \hline
\multirow{8}{*}{\rotatebox{90}{\textbf{Interactions}}} 
    & Fine-ideology interaction  &              &               &            &           & -0.110*   & -0.110*   \\
    &                            &              &               &            &           & (0.0618)  & (0.0618)  \\
    & Nudge-ideology interaction &              &               &            &           & -0.0266   & -0.0265   \\
    &                            &              &               &            &           & (0.0447)  & (0.0448)  \\
    & Fine-age interaction       &              &               & -0.0505*** &           &           &      \\
    &                            &              &               & (0.0112)   &           &           &      \\
    & Nudge-age interaction      &              &               & 0.0309**   &           &           &      \\
    &                            &              &               & (0.0130)   &           &           &      \\ \hline
    & Other Controls$^{a}$       & No           & Yes           & Yes        & Yes       & Yes       & Yes       \\ \hline
    & Constant                   & 1.456***     & -2.051        & -2.537     & -3.062   & -2.213    & -2.415    \\
    &                            & (0.312)      & (1.670)       & (1.721)    & (1.928)   & (1.666)   & (1.700)   \\
    & Observations               & 8,000        & 7,880         & 7,880      & 5,520 $^{b}$    & 7,880     & 7,880     \\
    & Clusters                   & 80           & 80            & 80         & 80        & 80        & 80     \\
\midrule
\multicolumn{8}{l}{\footnotesize{Robust standard errors in parentheses. Note that 6 subjects did not complete the SVO task.}} \\  
\multicolumn{8}{l}{\footnotesize{\shortstack[l]{ $^{a}$ Dummies for: religion, out of labor force, unemployed, federal regions, stay at home order in place. Continuous variables for: population \\ density, cumulative COVID-19 cases in state, daily COVID-19 deaths in state, years of schooling, and number of quiz failures. Specification 5\\also includes interactions between the Conservative Index and fine and nudge treatment variables.  } }} \\ 
\multicolumn{8}{l}{\footnotesize{\shortstack[l]{ $^{b}$ 276 subjects used in this regression. 124 subjects are excluded because they self-identified as ``Independent'' or ``Other'' with respect to\\political party support.} }} \\ 
\bottomrule
\end{tabular}
\end{table} }

\restoregeometry

\clearpage
\bibliographystyle{chicago}
\bibliography{main_bib.bib}

\renewcommand{\thefigure}{A\arabic{figure}}
\renewcommand{\thetable}{A\arabic{table}}
\setcounter{figure}{0}

\clearpage
\appendix
\singlespacing

\section{Online Appendix}
This Appendix contains additional information. Section \ref{sec:model} presents the theoretical framework and derives the hypotheses that we investigate in the experiment. Section \ref{sec:design} describes our dataset. Section \ref{sec:data_quality} summarizes our data quality protocols. Section \ref{sec:analysis} contains details of the analysis.

\subsection{Theory} \label{sec:model}
In this section, we set out the theoretical model with greater generality. The model is similar to those found in the theoretical economics literature on self-protection against a contagious process on fixed networks \citeplatex{CDG_2017, AMO_2016}.

Consider a set of $N={1,2,\ldots,n}$ risk-neutral agents on an unweighted and undirected network $\mathbf{G}$. If a link between agents $i$ and $j$ is present, then $G_{ij}=G_{ji}=1$. Otherwise $G_{ij}=G_{ji}=0$. Each agent $i$ simultaneously chooses whether to practice social distancing, at a cost $c>0$ to herself. 

After agents make this choice, exactly one agent is chosen uniformly at random to be exposed to COVID-19. Call this agent \emph{patient zero}. If patient zero is practicing social distancing, then she becomes infected with probability $\gamma<1$, and cannot pass COVID-19 on to anyone else. Otherwise, she becomes infected with probability 1 and passes it on to each of her neighbors who are also not practicing social distancing uniformly at random with probability $\alpha\in[0,1]$. 
Any other agent $j$ who becomes infected with COVID-19 passes it on to each of her own neighbors who are not practicing social distancing with probability $\alpha\in[0,1]$, again uniformly at random. An agent does not pass it on to any neighbor who is practicing social distancing. Therefore, an agent who practices social distancing may only become infected if she is patient zero.

At the end of the game, an agent receives a benefit $b>c$ if she is not infected, and 0 otherwise. Additionally, an agent pays a fine, $f \geq 0$ if she did not practice social distancing -- regardless of whether she infected anyone else or became infected herself. Note that setting $f = 0$ means no fine is present. When a subset of agents $S \subseteq N$ practice social distancing, the probability that agent $i$ becomes infected is $p_{i|S_i}$. This depends on the network structure and on who practices social distancing. Notice that if $i\in S$ then $p_{i|S_i}=\gamma / n$, since she can only become infected if she is patient zero.

We assume that all agents are self-interested -- they care only about themselves and so ignore any effects their choices have on others. Then agent $i$ receives an expected payoff, $\pi_i$, of:
\begin{align}
    \pi_{i} =
        \begin{cases}
        (1- \frac{\gamma}{n})b-c,   & \text{if } i \in S \\
        (1 - p_{i|S_i}) \cdot b - f,              & \text{otherwise.}
        \end{cases}
\end{align}

\noindent \textbf{Experimental algorithm.} The contagion process described above is implemented in the experiment as follows. Suppose agent $i$ is patient-zero who does not practice social distancing, and so is infected for sure. Then, $i$ interacts once with her first-order neighbors who do not practice social distancing. We record all neighbors of $i$ that become infected as a result of this interaction, and add them to a set $\text{Inf}(1)$. Next, each agent in set $\text{Inf}(1)$ interacts with her healthy first-order neighbors who do not practice social distancing. Any newly infected agents are added to a set $\text{Inf}(2)$, and in turn agents from this set interact with their healthy first-order neighbors who do not practice distancing in the second round of contagion. The process ends when no more agents are infected as a result of the last contagion step. After this, payoffs are calculated based on the decisions and health outcomes of individual players.\\

\noindent \textbf{Parameterization.} We now set out some of the parameter values and the networks we use in the experiment. Doing so allows us to state the hypotheses clearly. We have two different networks: a 5-node complete network, and a 5-node star network. In the complete network, each agent is connected to every other agent. In the star network, one agent is connected to every other agent, and there is no other link in the network. We call agents in the complete network ``close-knit'', agents at the center of the star ``superspreaders'', and agents on the arms of the star ``peripheral''. Denote these agents $C$, $S$, and $P$ respectively. 

\begin{figure}[t]
    \centering
    \includegraphics[width=\textwidth]{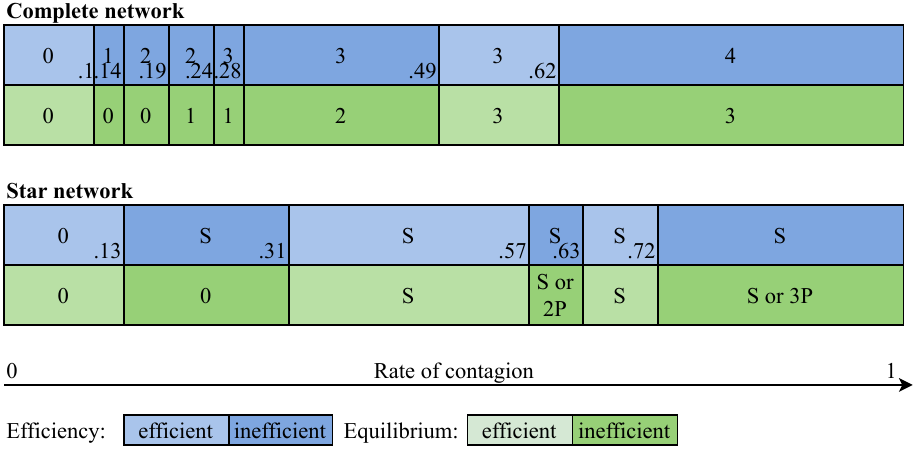}
    \caption{Equilibrium and socially efficient outcomes for the rate of contagion $\alpha \in [0,1]$. The other parameters are set at: $b=100$, $c=35$, $\gamma=0.5$, $N=5$. For the complete network, digits indicate the number of close-knit agents practicing social distancing in an equilibrium/efficient outcome. For the star network, S stands for the superspreader, and \#P indicates the number of peripheral agents practicing social distancing in an equilibrium/efficient outcome.}
    \label{fig:equilibrium}
\end{figure}

We set the cost $c$ of practicing social distancing equal to 35 points, the benefit $b$ of avoiding infection equal to 100 points, and the probability $\gamma$ of patient zero becoming infected if she is practicing social distancing equal to $0.5$. This allows us to investigate how the Nash equilibrium predictions and efficiency vary as the rate of contagion $\alpha$ is varied on $[0,1]$. Figure \ref{fig:equilibrium} shows the equilibrium and socially efficient sets in the complete network and the star network for all values of $\alpha$. A set is socially efficient if it maximizes the sum of the expected payoffs. In the experiment, we focus on two different levels of contagion: $\alpha=0.15$ (low contagion), and $\alpha=0.65$ (high contagion).

Below, we formulate a set of hypotheses that are the focus of our analysis. First, fines make distancing relatively more attractive and so we expect that they weakly increase the amount of social distancing in society, while nudges have no impact on payoffs. So trivially fines have a weakly greater effect than nudges. Note that due to the non-uniqueness of equilibria, it is possible that \emph{different} agents practice social distancing when fines are present. 

\begin{theor}\label{theor:fine_intervention}
    A fine $f>0$ for agents who do not practice social distancing weakly increases the number of agents practicing social distancing.
\end{theor}

\begin{theor}\label{theor:nudge_intervention}
    A behavioral `nudge', in the form of an informational video, has no impact on the social distancing decisions of any agent.
\end{theor}

\begin{theor}\label{theor:fine_vs_nudge}
    A fine $f>0$ increases the amount of social distancing weakly more than a behavioral `nudge'.
\end{theor}

The model predicts that the network position and contagiousness play important roles in determining social distancing decisions. Peripheral agents have few links and so a low incentive to practice social distancing. This means superspreaders ought to practice more distancing than close-knit agents, as superspreaders cannot rely on protection from others. All else equal, higher contagiousness makes infection more likely, increasing the incentive to practice distancing.

\begin{theor}\label{theor:position_effect}
    Superspreaders practice more social distancing than close-knit agents, and in turn, close-knit agents practice more social distancing than peripheral agents.
\end{theor}

\begin{theor}\label{theor:contagion_effect}
    Conditional on her position in the network, an agent practices more social distancing in the high contagion setting than in the low contagion setting.
\end{theor}

Regardless of their position in the network and the level of contagion, self-interested agents do not account for the benefits they provide to others -- in the form of reduced infection risk -- when they practice social distancing. Therefore, we expect that agents practice less social distancing than would be optimal. \Cref{theor:complete_network} confirms this intuition for the complete network, while \Cref{theor:star_network} shows that there can be too little or the right amount of social distancing (relative to the social optimum) in the star. Moreover, for certain other values of $\alpha$, \emph{too much} social distancing is also possible, due to the presence of multiple equilibria in the star network.

\begin{theor}\label{theor:complete_network}
    In the complete network, the Nash equilibrium involves fewer agents practicing social distancing than the social optimum.
\end{theor}

\begin{theor}\label{theor:star_network}
    In the star network: (1) in the low contagiousness environment, the Nash equilibrium involves fewer agents practicing social distancing than the social optimum; (2) in the high contagiousness environment, the Nash equilibrium and the social optimum coincide (with the hub agent practicing social distancing). 
\end{theor}

Practicing social distancing reduces the variability of an agent's own payoffs (as she is definitely healthy unless randomly selected to be patient zero) and also has external benefits for everyone else in the network (as it reduces the probability that any other agent becomes infected). Therefore more risk-averse agents and more altruistic preferences weakly increase the amount of social distancing an agent does. This does not depend on how we choose to model risk preferences and altruistic preferences.

\begin{theor}\label{theor:risk_aversion}
    The amount of social distancing an agent practices is weakly increasing in her level of risk aversion.
\end{theor}

\begin{theor}\label{theor:social_values}
    An agent with other-regarding (i.e. altruistic) preferences practices weakly more social distancing than in the Nash equilibrium.
\end{theor}

\subsection{Dataset} \label{sec:design}
This section supplements Section \ref{sec:data} and provides additional details on our dataset. Apart from data on subjects' decisions in the experiment, we collect data on a set of variables, which can be broadly categorized as follows: demographic controls, preference controls, and location-based controls. \Cref{tab:controls} presents summary statistics for some of these controls.

\begin{table}[t]
\centering
\begin{threeparttable}
\caption{Summary statistics the main controls.}
\label{tab:controls}
\begin{tabular}{lladl}
\toprule
Variable & & 
\multicolumn{1}{c}{$\bar{X}$} & 
\multicolumn{1}{c}{s.d.} & 
Comments \\ \midrule
\multicolumn{5}{l}{\underline{Demographic controls}} \\
Age & &  43.91 & 14.13 & Measured in years  \\
Gender & & 0.47 & 0.02 & Female = 1 \\
Race & & 0.84 & 0.02 & White = 1 \\
Education & & 15.34 & 2.04 & Measured in years \\
Out of labor force & & 0.19 & 0.02 & Yes = 1 \\
Unemployed & & 0.11 & 0.02 & Yes = 1 \\ 
Christian religion & & 0.39 & 0.02 & Yes = 1 \\
Other religion & & 0.12 & 0.02 & Yes = 1 \\
\multicolumn{5}{l}{\underline{Preference controls}} \\
Protect others & & 0.57 & 0.02 & Yes = 1, see text for details\\
Ideology score & & 3.08 & 3.06 & $\in [0,12]$, see text for construction details\\ 
Global warming skepticism score & & 1.99 & 3.10 & $\in [0,12]$, see text for construction details\\ 
BRET score & & 34.73 & 18.23 & $\in [0,100]$ \\
SVO type & & 0.58 & 0.02 & Prosocial = 1 \\ 
\multicolumn{5}{l}{\underline{Location-based controls}} \\
Population density & & 2.35 & 6.32 & Measured in 1,000's at county level \\
Cumulative cases & & 63.43 & 86.33 & Measured in 1,000's at state level \\
Daily deaths & & 54.81 & 64.23 & Measured at state level \\
Stay-at-home-orders & & 0.62 & 0.02 & Yes = 1, measured at state level \\ 
\bottomrule
\end{tabular}
\begin{tablenotes}
      \item Sample size is 400, except for SVO task which 6 subjects did not complete; $\bar{X}$ -- mean value, or proportion in case of binary variables; s.d. -- standard deviation. 
    \end{tablenotes}
\end{threeparttable}

\end{table}

\textbf{Demographic controls.} All participants are residents of the US, 47\% are female, and the mean age is just under 44 years. 84\% of our subjects are white, and 70\% are employed (either full- or part-time). Further, we estimate the number of years of education received based on educational attainment.\footnote{To do this we assume that all subjects took the standard number of years to complete each qualification, and undertook no education that did not lead to a qualification.} Based on this, the average subject in our sample has 15.3 years of education. Finally, 51\% of our participants consider themselves religious, with 39\% reporting Christianity as their religion.

\textbf{Preference controls.} A total of 92.5\% of subjects reported that they try to stay at home as much as possible because of the COVID-19 pandemic. Of these, 61.6\% (or 57\% of the full sample) report that the desire to `protect others' is one of the main reasons behind this decision. 

To capture subjects' political ideology, we construct an index from three questions in the post-experimental survey. These questions ask about: 1) subjects' support for President Donald Trump's handling of the COVID-19 pandemic, 2) their support for universal healthcare, and 3) their belief that social distancing measures impose unjustified economic costs. All are on a 5-point Likert scale \citeplatex{likert1932technique}, and so yield a 13-point index (0-12). Responses that indicate support for President Trump, opposition to universal healthcare, and belief in the unjustified economic costs of social distancing measures are scored positively. Higher scores, therefore, indicate a more Conservative ideology. In our sample, the mean ideology score is 3.1 (s.d. 3.1), and over 75\% of subjects have a score of 5 or less.

We also collect information on participants' attitudes to climate change. Specifically, our post-experimental survey includes three questions asking whether subjects believe that global warming is (1) happening, (2) caused mostly by human activity, and (3) affecting weather in the United States \citeplatex{HMML_2015a}. The resulting climate change attitudes index is on a 0-12 scale, with higher scores indicating greater skepticism towards climate change. The mean score in our sample is 1.99.

\begin{figure}[t]
    \centering
    \includegraphics[width=\textwidth]{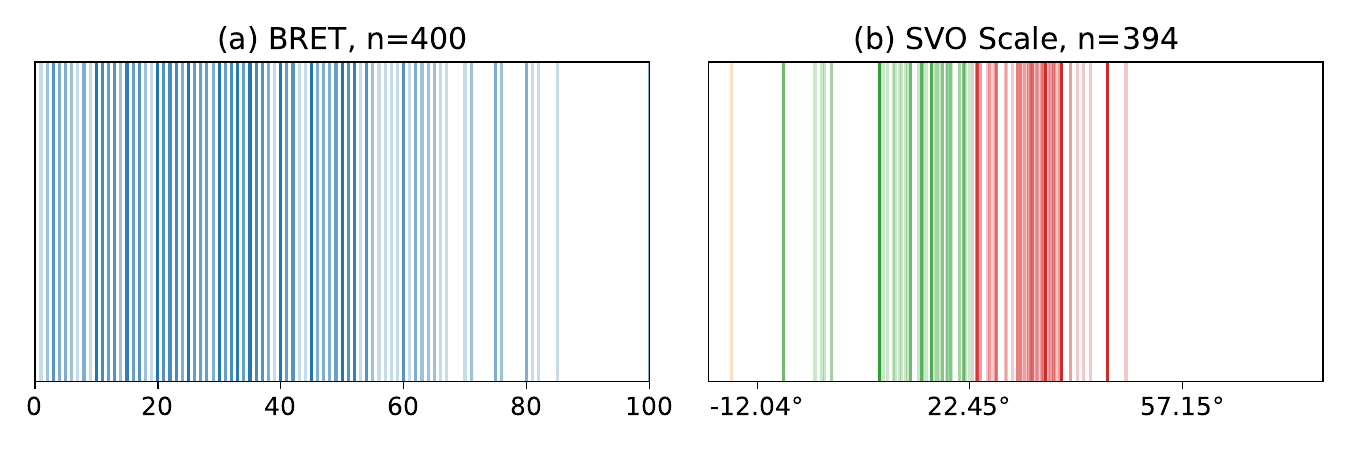}
    \caption{Bomb Risk Elicitation Task (BRET) and Social Value Orientation (SVO) Scale distributions for the sample. We draw a vertical line on the subplots for each subject whose score in BRET/SVO is of the corresponding value. More intense line color indicates that more subjects are concentrated at that value.}
    \label{fig:distribution_bret_svo}
\end{figure}

Further, we collect information on subjects' social value (SVO) and risk (BRET) preferences. \Cref{fig:distribution_bret_svo} presents the distributions of these preferences for our sample.  The average subject in the sample is moderately risk-averse, with a BRET score of 34.73 boxes. When it comes to social values, the majority of our subjects (58\%) are classified as prosocial (with SVO angle $\geq 22.45\degree$ and $<57.15\degree$). The second-largest category in the sample are individualists (the angle is $\geq 12.04\degree$ and $<22.45\degree$). Notice that we only have 1 subject who is classified as competitive (angle $<-12.04\degree$), and no subjects classified as altruistic (angle $\geq57.1\degree$).

\textbf{Location-based controls.} As part of both recruitment and the experiment, we collect subjects' IP-addresses. We use these to infer their geographic location, down to the county-level. \Cref{fig:distribution_location} shows the region-level distribution of our sample.

Further, using location information, we create a set of location-based control variables. First, for each subject, we record population density at the county level. The average density in the sample is 2.4k per square mile ~\citeplatex{census_2011,census_2019a}. Second, we use location data together with data on the development of the COVID-19 pandemic in the US. In particular, for each subject, we record the number of cumulative COVID-19 cases (average 63.4k) and new COVID-19 related deaths (average 59) in the state on the day of the experiment. We also record

\begin{table}[H]
\centering
\begin{threeparttable}
\caption{Age-gender-location distribution of the US adult population.}
\label{tab:population_distribution}
\begin{tabular}{@{}rl|adadadadadad@{}}
\toprule
\multirow{2}{*}{Region} & 
&\multicolumn{2}{c}{18-25} &\multicolumn{2}{c}{25-30} 
&\multicolumn{2}{c}{30-35} &\multicolumn{2}{c}{35-45} 
&\multicolumn{2}{c}{45-55} &\multicolumn{2}{c}{55+} 
\\
&  
&\multicolumn{1}{c}{M} &\multicolumn{1}{c}{F}  
&\multicolumn{1}{c}{M} &\multicolumn{1}{c}{F}  
&\multicolumn{1}{c}{M} &\multicolumn{1}{c}{F}  
&\multicolumn{1}{c}{M} &\multicolumn{1}{c}{F}  
&\multicolumn{1}{c}{M} &\multicolumn{1}{c}{F}  
&\multicolumn{1}{c}{M} &\multicolumn{1}{c}{F}  
\\
\midrule
I & & 0.29 & 0.29 & 0.20 & 0.19 & 0.19 & 0.19 & 0.34 & 0.35 & 0.38 & 0.41 & 0.85 & 1.00 \\
II & & 0.51 & 0.50 & 0.41 & 0.40 & 0.39 & 0.39 & 0.70 & 0.71 & 0.73 & 0.77 & 1.51 & 1.83 \\
III & & 0.56 & 0.54 & 0.43 & 0.42 & 0.41 & 0.41 & 0.74 & 0.76 & 0.78 & 0.81 & 1.70 & 2.01 \\
IV & & 1.21 & 1.15 & 0.92 & 0.91 & 0.83 & 0.84 & 1.58 & 1.65 & 1.66 & 1.74 & 3.64 & 4.33 \\
V & & 0.99 & 0.95 & 0.73 & 0.70 & 0.67 & 0.66 & 1.27 & 1.27 & 1.31 & 1.33 & 2.88 & 3.33 \\
VI & & 0.83 & 0.78 & 0.63 & 0.61 & 0.60 & 0.58 & 1.10 & 1.10 & 1.01 & 1.03 & 1.97 & 2.29 \\
VII & & 0.28 & 0.26 & 0.19 & 0.18 & 0.18 & 0.18 & 0.34 & 0.34 & 0.33 & 0.33 & 0.77 & 0.89 \\
VIII & & 0.25 & 0.23 & 0.19 & 0.18 & 0.18 & 0.17 & 0.33 & 0.31 & 0.28 & 0.27 & 0.60 & 0.65 \\
IX & & 0.98 & 0.92 & 0.81 & 0.76 & 0.75 & 0.71 & 1.34 & 1.31 & 1.27 & 1.28 & 2.54 & 2.92 \\
X & & 0.26 & 0.24 & 0.22 & 0.2 & 0.21 & 0.20 & 0.38 & 0.36 & 0.34 & 0.34 & 0.76 & 0.85 \\
\midrule
\multicolumn{13}{r}{\textbf{Total:}} & 100\% \\
\bottomrule
\end{tabular}
\begin{tablenotes}
      \item All numbers are in percentage terms; `18-25' etc. -- age categories; `M' -- males, `F' -- females; we divide states into 10 Standard Federal Regions (Office of Management and Budget) defined as follows: I -- CT, ME, MA, NH, RI, VT; II -- NJ, NY; III -- DE, DC, MD, PA, VA, WV; IV -- AL, FL, GA, KY, MS, NC, SC, TN; V -- IL, IN, MI, MN, OH, WI; VI -- AR, LA, NM, OK, TX; VII -- IA, KS, MO, NE; VIII -- CO, MT, ND, SD, UT, WY; IX -- AZ, CA, HI, NV; X -- AK, ID, OR, WA. Note that we exclude US territories.
    \end{tablenotes}
\end{threeparttable}

\vspace*{1cm}  

\includegraphics[width=\linewidth]{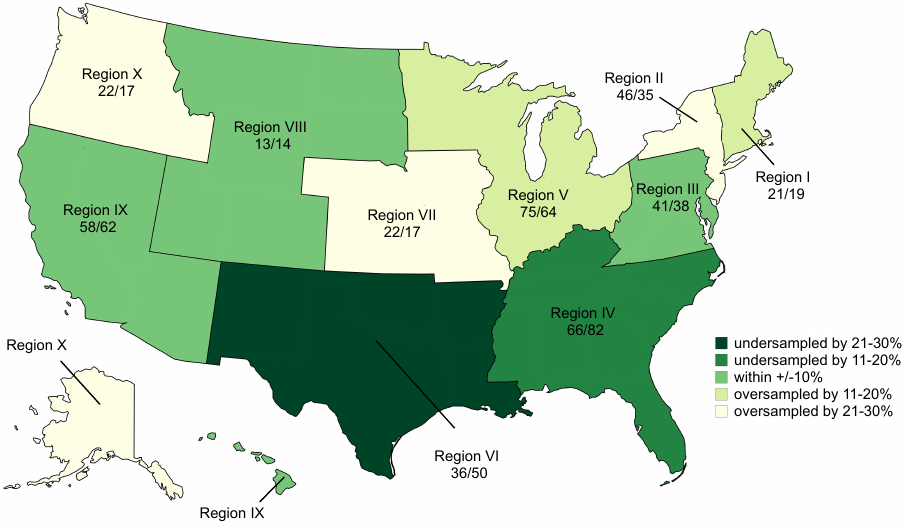}
\captionof{figure}{Location distribution, created with www.mapchart.net. States are divided into 10 Standard Federal Regions, using the same definition as \Cref{tab:population_distribution}. X/Y indicate participants counts, where X indicates the actual count in our sample, and Y -- the target representative count.}
\label{fig:distribution_location}
\end{table}

\noindent whether stay-at-home orders were in place in the participant's state on the day of the experiment. In our sample, 62\% of the subjects had stay-at-home orders in place on the day they participated in our experiment. 

\textbf{Sample representativeness.} We check the representativeness of our sample along the three dimensions -- i.e. age, gender, location. \Cref{tab:population_distribution} presents the distribution of the US adult population. The gender distribution of our sample is not statistically different from that of the adult US population (two-sided t-test, $p=0.11$). Further, \Cref{fig:distribution_location} presents a map of the US divided into the 10 Standard Federal Region. For each region, we indicate the target number of people and the realized count in our sample. The difference between the observed and target location distributions is not significant (two-sided $\chi^2$, $p=0.08$). When it comes to age, our distribution is unfortunately not quite representative of the adult US distribution (two-sided $\chi^2$, $p<0.0001$). In particular, the 35-45 and 45-55 age categories are over-represented, while the 55+ is under-represented.

\textbf{Sample balance.} We check that our sample is balanced with respect to treatment assignment along the three dimensions of interest -- gender, age category and geographical region of residence. We run a Pearson chi-squared test on participants' treatment assignment (an 8-category variable) against each of these three variables separately. All tests are not significant at the 10\% level. So we fail to reject the null hypothesis of no difference between our 8 treatments with respect to the three demographic characteristics. \Cref{tab:treatment_balance} reports the results of these tests.

\begin{table}[H]
\centering
\begin{threeparttable}
\caption{Pearson Chi-squared tests for balance of treatment assignment.}
\label{tab:treatment_balance}
\begin{tabular}{@{}llllllllll@{}}
\toprule
Variable & & & & & & & $\chi^2$ & p-val & df \\ \midrule
Gender (Female = 1) & & & & & & & 6.58 & 0.4735 & 7 \\
Age (6 categories as in \Cref{tab:population_distribution}) & & & & & & & 44.03 & 0.1409 & 35 \\
Region (10 categories as in \Cref{tab:population_distribution}) & & & & & & & 52.73 & 0.8185 & 63 \\ \bottomrule
\end{tabular}
\begin{tablenotes}
      \item $\chi^2$ -- value of the test statistic; p-val -- corresponding p-value, df -- degrees of freedom (number of distinct categories minus the number of parameters). 
    \end{tablenotes}
\end{threeparttable}
\end{table}

\subsection{Data quality protocols} \label{sec:data_quality}
Ensuring data quality is important, particularly in online experiments where experimenters have relatively less control of the environment. This section explains our protocols for ensuring data quality. 

\textbf{Recruitment.} We only make our survey visible to US residents who have completed at least 500 Human Intelligence Tasks (HITs) on MTurk \emph{and} have an approval rate of at least 96\%.\footnote{This means that the task-setters judged that they completed the tasks as required in at least 96\% of submitted tasks.} We keep track of IP addresses of participants and ensure that no two persons with the same IP address take part in the experiment. Multiple participation is not allowed even by participants who have only seen the instructions. 

Our recruitment survey also includes an attention check, which was failed by about 3\% of our standing panel. Existing research suggests that excluding subjects based on failed attention checks may not necessarily be appropriate \citeplatex{goodman2013data}. Therefore, we do not exclude subjects based on this attention check and rather use it as a robustness check for our analysis. In practice, 13 out of 400 subjects (a rate similar to that of the overall standing panel) in our dataset failed this check and excluding their data from the analysis does not affect the results reported in this paper.

\textbf{Experiment.} There are two main elements of our data quality control protocol at the experimental stage -- comprehension checks and the use of timers. First, subjects are quizzed on their understanding of instructions (both Baseline and Intervention). For the Baseline, they need to answer three questions concerning their understanding of the contagious mechanism and the way points for their actions are calculated. Subjects have three attempts and need to answer all three questions correctly to participate in the study. Each failed attempt is followed by an explanation of correct answers, and questions are different between attempts. After the instructions for the Intervention part of the experiment, subjects need to answer one question -- either asking how the fine works or what the video is about -- and have three attempts. These procedures are standard for all experiments in economics, both lab and online. We use the number of attempts subjects took to complete the instructions as a control. This variable is not significant at all conventional levels (see \Cref{tab:logit_main}). 

Next, throughout the experiment -- both Baseline and Intervention parts -- in each round subjects are given 20 seconds to make their social distancing decision. If a subject fails to submit the decision in time, she receives a penalty of 200 points and her decision is automatically recorded as a ``No''. A participant who fails to submit her decisions in three consecutive rounds is disqualified from the experiment and does not receive any compensation. A subject who is disqualified is replaced by a bot with pre-specified behavior (we use 4 different types of behavior) which allows the other members of her group to complete the experiment. Note that if the dropout occurs in Baseline, then the remaining subjects do not participate in the second part of the experiment, and just proceed to the post-experimental survey.

In practice, 6 subjects failed to submit their choices in three consecutive rounds of the Baseline. We had no instances of disqualification of this type in the Intervention part. We do not use data from the corresponding 6 groups in our analysis, but incorporating it back does not affect the results in robustness checks. The dataset analyzed in the paper contains a total of 13 missed choices across the 400 participants (0.08\% of the total number of decisions), which has no material impact on the results reported in this paper.

Finally, another potential stage where subjects can lose attention and drop out is between the Baseline and Intervention stages of our experiment. We cap waiting time after completing the quiz for the Intervention part of instructions to 10 minutes. That is, if the longest waiting subject in a group has been waiting for 10 minutes or more, the whole group skips the Intervention part and proceeds to post-experimental survey to avoid excessive waiting. In practice, we discarded 2 groups as a result of this `bottleneck' -- one in the fine treatment and another in the nudge. Again, the data from these groups is not part of our main analysis, but we perform a robustness check to ensure that our results are robust to using it.

\textbf{Post-experimental survey.} We include an attention check into our post-experimental questionnaire which was failed by 4 of the 400 subjects in our dataset. Again, we do not discard those subject from our main analysis, but dropping their data as part of robustness checks does not alter our results.

\subsection{Statistical analysis} \label{sec:analysis}
This section presents a detailed statistical analysis. Section \ref{sec:analysis_structural} tests for a structural break in the data between Baseline and Intervention. Section \ref{sec:analysis_aggregate} then presents analysis at the aggregate (group) level, and Section \ref{sec:analysis_individual} focuses on the individual level. Next, Section \ref{sec:robustness_convergence} contains convergence analysis for individual behavior, while Section \ref{sec:analysis_perception} reports on the difference between attitudes to interventions in the sample and observed behavior.

\subsubsection{Structural break identification} \label{sec:analysis_structural}

We use a Wald test to look for a single structural break in social distancing decisions over time. This involves performing a simple linear regression of social distancing decisions (left-hand side) on the round number (right-hand side), and then testing whether the coefficient on the time variable exhibits a structural break. The test assumes there is at most one structural break but is agnostic as to at what point it may occur.   

We perform this test on the data at three levels of aggregation. First, we consider all observations together. Second, we split by the rate of contagion and the policy intervention. This gives four equal-sized group: (1) $15\%$ contagion, fine; (2) $15\%$ contagion, nudge; (3) $65\%$ contagion, fine; and (4) $65\%$ contagion, nudge. Finally, we subdivide each of the four groups above by network type (i.e. complete vs star).

If the policy intervention is effective then we expect the test would find a structural break at round 21. With all observations together, the Wald test finds the structural break at round 21 ($p < 0.001$). When considering the dis-aggregated data, the Wald test finds a structural break at round 21 for each of the subsets, except for complete networks with the nudge treatment.

\begin{table}[ht]
\centering
\caption{Results from Structural Break analysis using the Wald Test}\label{tab:structural break}
\begin{tabular}{c|c}
\textbf{Low contagion: $\alpha = 0.15$} & \textbf{High contagion $\alpha = 0.65$} \\
\hline
\begin{tabular}{llll}
\textbf{Treatment}    & \textbf{Network}  & \textbf{Break}  & \textbf{p-value} \\
fine   & all               & $21$                  & $< 0.001$        \\
fine   & complete          & $21$                  & $< 0.001$       \\
fine   & star              & $21$                  & $< 0.001$       \\
nudge  & all               & $21$                  & $< 0.001$        \\
nudge  & complete          & $7$                   & $< 0.001$       \\
nudge  & star              & $21$                  & $< 0.001$       \\
\end{tabular}
&
\begin{tabular}{llll}
\textbf{Treatment}    & \textbf{Network}  & \textbf{Break}  & \textbf{p-value} \\
fine   & all               & $21$                  & $< 0.001$        \\
fine   & complete          & $21$                  & $< 0.001$       \\
fine   & star              & $21$                  & $< 0.001$       \\
nudge  & all               & $21$                  & $< 0.001$        \\
nudge  & complete          & $16$                  & $< 0.001$       \\
nudge  & star              & $21$                  & $0.009$       \\
\end{tabular}

\end{tabular}
\end{table}

\subsubsection{Aggregate level} 
\label{sec:analysis_aggregate}

In this section, we focus on the aggregate analysis of the decision data and conduct a set of nonparametric tests. Since individual observations are correlated at the group level, our unit of observation is a group of 5 participants. Recall that we have 80 groups in total, equally distributed across 8 treatments. For each group, we calculate the average proportion of participants that practice social distancing in the last 10 rounds of both parts of the experiment. Below, we refer to this average as `distancing levels'. We discard the first 10 rounds of both parts of the experiment because our participants display clear convergence behavior. In particular, by round 11, in both Baseline and Intervention, at least 80\% of participants converge to a particular strategy in all treatments. Further details on convergence behavior is in Section \ref{sec:robustness_convergence}. All key results below are also robust to including all 20 rounds of both parts of the experiment. 

\Cref{tab:nonparametric_analysis} presents the results of the non-parametric tests. All of the findings of this section are robust to using parametric analogs of these non-parametric tests. In our aggregate analysis, we test the set of hypotheses in Section \ref{sec:model}. Note that we cannot test \Cref{theor:risk_aversion,,theor:social_values} here, which are instead covered in Section \ref{sec:analysis_individual}. 

Part 1 of \Cref{tab:nonparametric_analysis} shows that the data from the Baseline from fine and nudge treatments can be pulled together as these samples are statistically identical at all conventional significance levels. This result is robust to various specifications.

\afterpage{\clearpage
\setlength\LTleft{-10mm}
\begin{ThreePartTable}
\begin{TableNotes}
  \item Null Hypothesis: for each hypothesis, sample \#1 and sample \#2 appear consecutively \textit{in italics}; Alt: either 2-sided or 1-sided alternative, a 1-sided alternative is always s.t. median(sample \#1) $>$ median(sample \#2); Test: Mann-Whitney U-test (MW) for unmatched samples and Wilcoxon Signed-Rank test (WSR) for matched samples; n: the number of observations per sample, where one observation is one group of 5 subjects; $\Delta \bar{X}$: difference in means between the two samples; $\Delta \tilde{X}$: difference in medians between the two samples; p-val: p-value for the test of the Null Hypothesis against the Alternative; Sig: significance of the test where * -- 10\%, ** -- 5\%, *** -- 1\%.
\end{TableNotes}
\begin{longtable}{lccccadada}
\caption{Non-parametric analysis at the aggregate level, last 10 rounds used.}
\label{tab:nonparametric_analysis}\\
\toprule
Null Hypothesis & \multicolumn{1}{c}{\begin{tabular}[c]{@{}c@{}}Rate of\\ contagion\end{tabular}} & \multicolumn{1}{c}{Position} & \multicolumn{1}{c}{Alt} & \multicolumn{1}{c}{Test} &  \multicolumn{1}{c}{n} & \multicolumn{1}{c}{$\Delta \bar X$ } &
\multicolumn{1}{c}{$\Delta \tilde X$ } & \multicolumn{1}{c}{p-val} & \multicolumn{1}{c}{Sig} \\ \midrule
\endfirsthead
\multicolumn{10}{c}{\textit{Continued from previous page}} \\
\toprule
Null Hypothesis & \multicolumn{1}{c}{\begin{tabular}[c]{@{}c@{}}Rate of\\ contagion\end{tabular}} & \multicolumn{1}{c}{Position} & \multicolumn{1}{c}{Alt} & \multicolumn{1}{c}{Test} &  \multicolumn{1}{c}{n} & \multicolumn{1}{c}{$\Delta \bar X$ } &
\multicolumn{1}{c}{$\Delta \tilde X$ } & \multicolumn{1}{c}{p-val} & \multicolumn{1}{c}{Sig} \\ \midrule
\endhead
\multicolumn{10}{r}{\textit{Continued on next page}} \\
\endfoot
\hline
\insertTableNotes 
\endlastfoot
\multicolumn{10}{l}{Part 1. Baseline} \\
\hline
\multirow{6}{3.5cm}{\textit{Fine} vs \textit{nudge}} & all & all & \multirow{6}{*}{2-sided} & \multirow{6}{*}{MW} & 40 & .03 & .05 & .33 & \\ & all & close-knit & & & 20 & .01 & 0 & .81 & \\ & all & superspreader & & & 20 & .02 & .15 & .47 & \\ & all & peripheral & & & 20 & .07 & .1 & .20 & \\ & 15\% & all & & & 20 & .02 & -.01 & .73 & \\ & 65\%  & all & & & 20 & .05 & .01 & .34 & \\
\multicolumn{10}{l}{Part 2. Intervention vs Baseline. \Cref{theor:fine_intervention} (fine has an effect)} \\
\hline
\multirow{6}{3.5cm}{Fine: \textit{Intervention} vs \textit{Baseline}} & all & all & \multirow{6}{*}{1-sided} & \multirow{6}{*}{WSR} & 40 & .07 & .03 & .0008 & *** \\ & all & close-knit & & & 20 & .07 & .07 & .01 & ** \\ & all & superspreader & & & 20 & .09 & 0 & .09 & * \\ & all & peripheral & & & 20 & .07 & 0 & .04 & ** \\ & 15\% & all & & & 20 & .05 & .03 & .04 & ** \\ & 65\% & all & & & 20 & .09 & .1 & .003 & *** \\
\hline
\multicolumn{10}{l}{Part 3. Intervention vs Baseline. \Cref{theor:nudge_intervention} (nudge has an effect)} \\
\hline
\multirow{6}{3.5cm}{Nudge: \textit{Intervention} vs \textit{Baseline}} & all & all & \multirow{6}{*}{1-sided} & \multirow{6}{*}{WSR} & 40 & .04 & .04 & .008 & *** \\ & all & close-knit & & & 20 & .04 & .07 & .05 & ** \\ & all & superspreader & & & 20 & .01 & .05 & .50 & \\ & all & peripheral & & & 20 & .04 & .06 & .06 & * \\ & 15\% & all & & & 20 & 0.0 & -0.1 & .50 & \\ & 65\% & all & & & 20 & .07 & .08 & .0003 & *** \\
\hline
\multicolumn{10}{l}{Part 4. Intervention. \Cref{theor:fine_vs_nudge} (fine is more effective than nudge)}\\
\hline
\multirow{6}{3.5cm}{\textit{Fine} vs \textit{nudge}} & all & all & \multirow{6}{*}{1-sided} & \multirow{6}{*}{MW} & 40 & .07 & .04 & .08 & * \\ & all & close-knit & & & 20 & .04 & 0 & .24 & \\ & all & superspreader & & & 20 & .1 & .1 & .06 & * \\ & all & peripheral & & & 20 & .09 & .04 & .14 & \\ & 15\% & all & & & 20 & .07 & .03 & .15 & \\ & 65\% & all & & & 20 & .07 & .03 & .12 & \\
\hline
\multicolumn{10}{l}{Part 5. Baseline. \Cref{theor:position_effect} (superspreader $>$ close-knit $>$ peripheral)}\\
\hline
\multirow{3}{3.5cm}{\textit{Superspreader} vs \textit{close-knit}} & all & -- & \multirow{3}{*}{1-sided} & \multirow{3}{*}{MW} & 40 & .1 & .19 & .004 & *** \\ & 15\% & -- & & & 20 & .1 & .18 & .03 & ** \\ & 65\% & -- & & & 20 & .1 & .21 & .03 & ** \\ \hline
\multirow{3}{3.5cm}{\textit{Close-knit} vs \textit{peripheral}} & all & -- & \multirow{3}{*}{1-sided} & \multirow{3}{*}{MW} & 40 & .16 & .11 & .0001 & *** \\ & 15\% & -- & & & 20 & .17 & .20 & .006 & *** \\ & 65\% & -- & & & 20 & .15 & .01 & .003 & *** \\ \hline
\multirow{3}{3.5cm}{\textit{Superspreader} vs \textit{peripheral}} & all & -- & \multirow{3}{*}{1-sided} & \multirow{3}{*}{WSR} & 40 & .26 & .30 & $<$.0001 & *** \\ & 15\% & -- & & & 20 & .27 & .38 & .0007 & *** \\ & 65\% & -- & & & 20 & .25 & .31 & .0001 & *** \\ 
\hline
\pagebreak
\multicolumn{10}{l}{Part 6. Intervention. \Cref{theor:position_effect} (superspreader $>$ close-knit $>$ peripheral)}\\
\hline
\multirow{3}{3.5cm}{\textit{Super-spreader} vs \textit{close-knit}} & all & -- & \multirow{3}{*}{1-sided} & \multirow{3}{*}{MW} & 40 & .1 & .17 & .002 & *** \\ & 15\% & -- & & & 20 & .12 & .16 & .01 & ** \\ & 65\% & -- & & & 20 & .08 &. 12 & .01 & ** \\ \hline
\multirow{3}{3.5cm}{\textit{Close-knit} vs \textit{peripheral}} & all & -- & \multirow{3}{*}{1-sided} & \multirow{3}{*}{MW} & 40 & .16 & .16 & .0002 & *** \\ & 15\% & -- & & & 20 & .21 & .18 & .0009 & *** \\ & 65\% & -- & & & 20 & .11 & .16 & .02 & ** \\ \hline
\multirow{3}{3.5cm}{\textit{Superspreader} vs \textit{peripheral}} & all & -- & \multirow{3}{*}{1-sided} & \multirow{3}{*}{WSR} & 40 & .25 & .33 & $<$.0001 & *** \\ & 15\% & -- & & & 20 & .32 & .34 & .0001 & *** \\ & 65\% & -- & & & 20 & .19 & .28 & .0007 & *** \\ 
\hline
\multicolumn{10}{l}{Part 7. Baseline. \Cref{theor:contagion_effect} (more distancing in high contagion compared to low contagion)}\\
\hline
\multirow{4}{3.5cm}{\textit{65\%} vs \textit{15\%} rate of contagion} & all & all & \multirow{4}{*}{1-sided} & \multirow{4}{*}{MW} & 40 & .1 & .06 & .01 & ** \\ & all & close-knit & & & 20 & .09 & .07 & .09 & * \\ & all & superspreader & & & 20 & .09 & .1 & .16 & \\ & all & peripheral & & & 20 & .11 & .16 & .03 & ** \\ 
\hline
\multicolumn{10}{l}{Part 8. Intervention. \Cref{theor:contagion_effect} (more distancing in high contagion compared to low contagion)}\\
\hline
\multirow{4}{3.5cm}{\textit{65\%} vs \textit{15\%} rate of contagion} & all & all & \multirow{4}{*}{1-sided} & \multirow{4}{*}{MW} & 40 & .15 & .14 & .0002 & *** \\ & all & close-knit & & & 20 & .12 & .14 & .009 & *** \\ & all & superspreader & & & 20 & .08 & .1 & .05 & ** \\ & all & peripheral & & & 20 & .22 & .16 & .002 & *** \\ 
\hline
\multicolumn{10}{l}{Part 9. Baseline. \Cref{theor:complete_network} (actual versus equilibrium/efficient outcomes in complete network)}\\
\hline
\multirow{2}{3.5cm}{\textit{Actual} vs \textit{equilibrium}} & 15\% & all & \multirow{2}{*}{1-sided} & \multirow{2}{*}{MW} & 20 & .63 & .62 & $<$.0001 & *** \\ & 65\% & all & & & 20 & .11 & .09 & .004 & *** \\ \hline
\multirow{2}{3.5cm}{\textit{Actual} vs \textit{efficient}} & 15\% & all & \multirow{2}{*}{1-sided} & \multirow{2}{*}{MW} & 20 & .23 & .22 & $<$.0001 & *** \\ & 65\% & all & & & 20 & -.09 & -.11 & .01 & ** \\
\hline
\multicolumn{10}{l}{Part 10. Baseline. \Cref{theor:star_network} (actual versus equilibrium/efficient outcomes in star network)}\\
\hline
\multirow{4}{3.5cm}{\textit{Actual} vs \textit{equilibrium}} & \multirow{2}{*}{15\%} & superspreader & \multirow{4}{*}{1-sided} & \multirow{4}{*}{MW} & 20 & .73 & .8 & $<$.0001 & *** \\ & & peripheral & & & 20 & .46 & .42 & $<$.0001 & *** \\ & \multirow{2}{*}{65\%} & superspreader & & & 20 & -.18 & -.1 & $<$.0001 & *** \\ & & peripheral & & & 20 & .57 & .59 & $<$.0001 & *** \\ \hline
\multirow{2}{3.5cm}{\textit{Actual} vs \textit{efficient}} & \multirow{2}{*}{15\%} & superspreader & \multirow{2}{*}{1-sided} & \multirow{2}{*}{MW} & 20 & -.27 & -.2 & $<$.0001 & *** \\ & & peripheral & & & 20 & .46 & .42 & $<$.0001 & *** \\ 
\bottomrule
\end{longtable}
\end{ThreePartTable}
\clearpage
}

In Part 2 of \Cref{tab:nonparametric_analysis}, we examine the effect of introducing a fine in Intervention. From the table, it is evident that a fine has a positive and statistically significant effect on distancing levels compared to the Baseline with no fine in all specifications considered. Most results in this part are statistically significant at least at the 5\% level. The effect of the fine is also sizable -- it results in an increase in mean distancing levels of 5\% to 9\% depending on the specification. The following result summarizes the above observations. Notice that the numbering of all results in this section corresponds to the numbering of Hypotheses in Section \ref{sec:model}.\\

\begin{findingA}
    The introduction of the fine increases the level of social distancing. The effect is observed for all three positions as well as for the two rates of contagion. The effect is both statistically significant (WSR, $p<0.05$ in all but one specification where $p=0.09$) and sizable in magnitude.
\end{findingA}

In Part 3 of \Cref{tab:nonparametric_analysis} we repeat the analysis from Part 2, but now looking at the nudge Intervention rather than the fine. The tests find a statistically significant effect of the nudge in only 4 of the 6 specifications. Generally, we can see that the nudge has a smaller effect which is not as robust as that of the fine. In particular, the effect of the nudge is very significant and large in magnitude (7\%) under high contagion, but disappears completely both statistically and in terms of economic magnitude under low contagion. Also, the nudge seems to have no effect on the superspreader. Note also that the effects of the nudge are not robust to using data from all rounds of Baseline and Intervention.

\begin{findingA}
    The introduction of the nudge generally increases the levels of social distancing. The effect, however, is not robust to alternative specifications and is smaller in magnitude than that of the fine. In particular, the effect is statistically significant (WSR, $p=0.0003$) and sizeable in magnitude under high contagion, but not low contagion (WSR, $p=0.5$). Further, a statistically significant effect is observed for the close-knit ($p=0.05$) and peripheral (WSR, $p=0.06$) participants but not for superspreaders.
\end{findingA}

Part 4 of \Cref{tab:nonparametric_analysis} shows that the data from the Intervention in the second part of the experiment is not statistically different for the fine and nudge treatments. In particular, even though the mean level of distancing in Intervention in fine treatments is higher than that in nudge treatments, our non-parametric testing fails to find any difference between them in 4 out of 6 specifications. We do, however, find evidence that the fine is more effective than the nudge (1) when data is pulled across treatments, and (2) for the superspreader. The effect is substantial in magnitude (7-10\%) and significant at 10\% level. 

\begin{findingA}
    There is limited evidence that the fine is more effective than the nudge, and the effect is not robust. In particular, the fine increases distancing levels by more than the nudge (1) when all data is pulled together (MW, $p=0.08$), and (2) separately for the superspreader (MW, $p=0.06$). The effect, however, is not statistically significant for other specifications.
\end{findingA}

In Part 5 of \Cref{tab:nonparametric_analysis}, we compare distancing levels in the three positions -- close-knit, superspreader, and peripheral -- in Baseline. From the table, we can see that (1) distancing levels are higher in the superspreader position relative to both the close-knit and the peripheral, and (2) higher in the close-knit relative to the peripheral. We run hypothesis tests both (a) aggregating over the rate of contagion, and (b) separately by two levels of the rate of contagion. All results in this part of the table are significant at the 5\% level, with most also being significant at 1\%. The differences between mean distancing levels in different positions are also large in magnitude -- between 10\% and 27\% depending on the specification. 

Part 6 of \Cref{tab:nonparametric_analysis} repeats the analysis in Part 5, but now focusing on Intervention, rather than Baseline. The conclusions here are the same as above, with all results being significant at the 5\% and most also at the 1\% level.

\begin{findingA}
    Superspreaders practice more social distancing than close-knit participants, who, in turn, practice more distancing than peripheral participants. The differences are both statistically significant (MW, $p<0.05$) and large in magnitude in all specifications. 
\end{findingA}

In Parts 7 and 8 of \Cref{tab:nonparametric_analysis}, we investigate the effects on the rate of contagion on distancing behavior, separately for Baseline and Intervention. We find that there is generally significantly more social distancing under 65\% rate of contagion relative to 15\%. The test is not statistically significant only for the superspreader in Baseline. The effect is also large in magnitude -- an average of 8-22\% depending on the specification. Note that the effect on superspreaders is smallest, but is probably explained, at least in part, by ceiling effects. In particular, in all treatments, the mean distancing levels in the last 10 rounds in the superspreader position are at least 70\%, so there is limited room for a further increase.

\begin{findingA}
    Subjects in the experiment generally practice more social distancing in high contagion environments relative to low contagion environments. The effect is statistically significant for close-knit (MW, $p=0.9$ in Baseline and $p=0.009$ in Intervention) and peripheral (MW, $p=0.03$ and $p=0.002$ in Baseline and Interventon respectively) subjects. For superspreaders, the effect is relatively smaller, and less robust to specifications (MW, $p=0.16$ in Baseline and $p=0.05$ in Intervention).
\end{findingA}

Finally, in Parts 9 and 10 of the table, we focus on testing \Cref{theor:complete_network,,theor:star_network}. Specifically, we compare the outcomes in the complete and the star networks to (1) predictions of the Nash equilibrium, and (2) efficient outcomes.

We find that the levels of social distancing are not in line with theoretical predictions, with all tests being statistically significant at least at the 5\% level. In particular, distancing levels in the complete network are well above those predicted by Nash equilibrium. The difference is particularly large in the low contagion environment, where the equilibrium prediction is no social distancing, but the actual average level of social distancing in the last 10 rounds of Baseline stands at 62\%. Distancing levels are also higher than those predicted by efficiency under low contagion. On the other hand, in the high contagion environment, the actual distancing levels are below efficiency requirements.

\begin{findingA}
     Prior to intervention, the observed levels of social distancing in the complete differ from equilibrium and efficiency predictions. In particular, more social distancing is observed than predicted by Nash equilibrium. When it comes to efficiency, actual levels of distancing are above efficient under low contagion, but below efficient under high contagion. All results are statistically significant (MW, $p\leq0.01$).
\end{findingA}

In the star network, the amount of social distancing done by the peripheral participants is well above both equilibrium predictions and efficiency requirements. On the other hand, the levels of social distancing observed for the superspreaders is above the equilibrium but below the efficient level for low contagion, and below both the equilibrium and efficient levels for high contagion.

\begin{findingA}
    Prior to intervention, the observed levels of social distancing in the star network differ from equilibrium and efficiency predictions. The result is true for both positions and both rates of contagion. In low contagion, participants in both positions practice more distancing than predicted by equilibrium analysis, while in high contagion superspreaders practice less and peripheral participants practice more than predicted. When it comes to efficiency, peripheral agents practice more social distancing and superspreaders practice less social distancing in both high and low contagion. All results are statistically significant (MW, $p<0.0001$).
\end{findingA}

\subsubsection{Individual level} \label{sec:analysis_individual}
We now turn to individual behavior, and examine the determinants of individual social distancing decisions. To do this, we use a Random Effects Logit model. This implicitly models the following Random Utility framework for binary choice:
\begin{align}
    y_{it} = 1 \iff x_{it} \beta + v_{i} + \epsilon_{it} > 0.
\end{align}

According to this model, subjects choose to practice social distancing if and only if they receive greater utility from doing so than from not doing so. Note that (implicitly) we have normalized the utility from not distancing to zero -- this is without loss of generality. This is a flexible and tractable way to model binary decisions. The model assumes that the subject-specific random effect, $v_{i}$, is normally distributed and that $\epsilon_{it}$ follows the logistic distribution. 

In addition, we only use data from the final 10 rounds of each part of the experiment, because it takes time for subjects' behavior to converge. We have five categories of controls that collectively cover a wide variety of factors. First, and most obviously, we control for the experimental treatments: the fine, nudge, rate of contagion, and network position (model 1 of \cref{tab:logit_main}). As we randomly assign subjects to these treatments, we can be confident that their effects are causal. 

Next, we add controls for social demographics (in model 2), geographic and institutional factors (in model 3), social and risk preferences (in model 4), and finally, ideology (in model 5). To complete our model, we also control for interactions between ideology and the policy interventions (fine/nudge). This is because we find that subjects' responsiveness to the fine depends on their ideology -- more conservative subjects are \emph{less responsive} to the fine. \\

\noindent \textbf{Results.} Ex ante, it seems plausible that each of the controls could be related to social distancing decisions. However, as we can see in model 6, only some of them are.\footnote{The p-values in all results relate to model 6.} All experimental treatments have a significant effect -- both statistically and in terms of economic relevance -- and their direction is intuitive. Statistical tests are t-test on coefficients in the Random Effects Logit regression (REL hereafter), or t-test on coefficients in the instrumental variables Random Effects Logit (IV, hereafter).

First, the fine significantly increases social distancing (\Cref{theor:fine_intervention}), while the nudge marginally increases social distancing (\Cref{theor:nudge_intervention}). Qualitatively, the fine is more effective (the statistical significance of the nudge is also not robust), but in our preferred model, the difference between the two effects is not statistically significant (\Cref{theor:fine_vs_nudge}). Note that the numbering of the results corresponds exactly to the number of the hypotheses in Section \ref{sec:model}, and the ``I'' prefix denotes individual-level results.

\begin{findingI}
    The fine increases the probability that an individual practices social distancing (REL, $p=0.001$).
\end{findingI}

\begin{findingI}
    The nudge marginally increases the probability that an individual practices social distancing (REL, $p=0.09$). The effect is approximately half that of the fine and is not robust to changes in the regression specification.
\end{findingI}

\begin{findingI}
    The effect of the fine is larger than the effect of the nudge, but the difference is not statistically significant (REL, $p=0.11$).
\end{findingI}

Second, superspreaders distance more than `close-knit' agents, even though they have the same number of links in the network (\Cref{theor:position_effect}). This suggests that when subjects know they are relatively highly connected, they take some action to compensate. Note that our experimental design cannot identify \emph{why} they do this -- whether to protect themselves or to protect others. Peripheral agents distance significantly less -- being less exposed to community contagion in the first place reduces the likelihood that agents take protective action. Further, subjects do more social distancing in a high contagion environment than in a low contagion environment, irrespective of their network position (\Cref{theor:contagion_effect}).

\begin{findingI}
    Superspreaders practice social distancing with a greater probability than close-knits subjects (REL, $p=0.05$), who in turn, practice social distancing with a greater probability than peripheral subjects (REL, $p<0.0001$). 
\end{findingI}

\begin{findingI}
    Subjects practice social distancing with a greater probability in the high contagion setting than in the low contagion setting (REL, $p<0.0001$).
\end{findingI}

Beyond the experimental treatments, we find significant effects for age, gender, race, social and risk preferences, and ideology. Older, female, and non-white subjects all practice more social distancing. More risk-averse agents and prosocial agents (as classified by the SVO task) also practice more social distancing (\Cref{theor:risk_aversion} and \Cref{theor:social_values}).

\stepcounter{findingI}

\stepcounter{findingI}

\begin{findingI}
    More risk-averse subjects practice social distancing with a higher probability (REL, $p=0.001$).
\end{findingI}

\begin{findingI}
    Subjects who indicate greater concern for others' well-being practice social distancing with a higher probability (REL, $p<0.0001$).
\end{findingI}

Perhaps the most interesting non-treatment effect is the subjects' political ideology. Subjects with stronger Conservative ideology practice less social distancing, and are marginally \emph{less responsive} to fines. Recall that our ideology index is constructed from subjects' responses to three questions from the post-experiment questionnaire. They ask about subjects' support for President Donald Trump's handling of the COVID-19 pandemic, their support for universal healthcare, and their belief that social distancing measures impose unjustified economic costs. 

\begin{findingI}
    More conservative subjects both practice distancing with a lower probability (iv, $p=0.04$) and are marginally less responsive to the fine (IV, $p=0.08$).
\end{findingI}

\afterpage{\clearpage
\setlength\LTleft{-17.5mm}
\begin{ThreePartTable}
\begin{TableNotes}
  \item The variable ``Quiz attempts'' measure the number of attempts subjects required to pass the quiz prior to the main experiment. It is a proxy for a subject's sophistication. 
  \item Population density = 1000's of people per square mile. Cumulative cases = 1000's of confirmed cases in state.
  \item Significance levels: * -- 10\%, ** -- 5\%, *** -- 1\%.
\end{TableNotes}
\begin{longtable}{ladadada}
\caption{Main logit regression results.}
\label{tab:logit_main} \\
\toprule
&  \multicolumn{6}{c}{Decision to practice distancing (binary)} \\ \cline{2-8} 
\multicolumn{1}{c}{Variables} & \multicolumn{1}{c}{(1)} & \multicolumn{1}{c}{(2)} & \multicolumn{1}{c}{(3)} & \multicolumn{1}{c}{(4)} & \multicolumn{1}{c}{(5)} & \multicolumn{1}{c}{\begin{tabular}[c]{@{}c@{}}(6)\\Main\end{tabular}} & \multicolumn{1}{c}{\begin{tabular}[c]{@{}c@{}}(7)\\IV\end{tabular}}\\ 
\midrule
\endfirsthead
\multicolumn{8}{c}{\textit{Continued from previous page}} \\
\toprule
&  \multicolumn{6}{c}{Decision to practice distancing (binary)} \\ \cline{2-8} 
\multicolumn{1}{c}{Variables} & \multicolumn{1}{c}{(1)} & \multicolumn{1}{c}{(2)} & \multicolumn{1}{c}{(3)} & \multicolumn{1}{c}{(4)} & \multicolumn{1}{c}{(5)} & \multicolumn{1}{c}{\begin{tabular}[c]{@{}c@{}}(6)\\Main\end{tabular}} & \multicolumn{1}{c}{\begin{tabular}[c]{@{}c@{}}(7)\\IV\end{tabular}}\\ 
\midrule
\endhead
\multicolumn{8}{r}{\textit{Continued on next page}} \\
\endfoot
\hline
\insertTableNotes 
\endlastfoot
\multicolumn{8}{l}{\textbf{\underline{Treatment}}} \\
Fine & 0.755*** & 0.752*** & 0.752*** & 0.762*** & 0.761*** & 1.146*** & 1.147*** \\
 & (0.231) & (0.234) & (0.234) & (0.242) & (0.242) & (0.352) & (0.352) \\
Nudge & 0.347** & 0.353** & 0.353** & 0.344** & 0.343** & 0.438* & 0.437* \\
 & (0.157) & (0.154) & (0.154) & (0.155) & (0.155) & (0.261) & (0.261) \\
High contagion & 1.330*** & 1.477*** & 1.404*** & 1.529*** & 1.537*** & 1.547*** & 1.543*** \\
 & (0.336) & (0.334) & (0.318) & (0.330) & (0.323) & (0.323) & (0.322) \\
Superspreader & 0.975** & 1.234*** & 1.184*** & 0.927** & 0.830** & 0.833** & 0.846** \\
 & (0.438) & (0.450) & (0.423) & (0.427) & (0.421) & (0.424) & (0.429) \\
Peripheral & -1.608*** & -1.364*** & -1.416*** & -1.684*** & -1.780*** & -1.785*** & -1.771*** \\
 & (0.322) & (0.332) & (0.319) & (0.334) & (0.333) & (0.333) & (0.340) \\
\multicolumn{8}{l}{\textbf{\underline{Demographic controls}}} \\
Age (years) &  & 0.0859*** & 0.0858*** & 0.0813*** & 0.0793*** & 0.0796*** & 0.0796*** \\
 &  & (0.0129) & (0.0135) & (0.0144) & (0.0137) & (0.0139) & (0.0139) \\
Female &  & 1.237*** & 1.193*** & 0.909** & 1.069*** & 1.072*** & 1.054*** \\
 &  & (0.352) & (0.367) & (0.374) & (0.372) & (0.372) & (0.380) \\
Race = white &  & -1.310*** & -1.553*** & -1.149** & -0.960** & -0.955** & -0.986** \\
 &  & (0.462) & (0.480) & (0.463) & (0.469) & (0.468) & (0.464) \\
Education (years) &  & 0.106 & 0.0490 & 0.0104 & 0.00341 & 0.00423 & 0.00713 \\
 &  & (0.0908) & (0.0916) & (0.0958) & (0.0984) & (0.0984) & (0.0977) \\
Religion = Christian &  & -0.517 & -0.443 & -0.287 & 0.0632 & 0.0616 & 0.00188 \\
 &  & (0.409) & (0.410) & (0.413) & (0.436) & (0.437) & (0.439) \\
Religion = other &  & -0.166 & -0.0445 & 0.278 & 0.384 & 0.370 & 0.347 \\
 &  & (0.543) & (0.560) & (0.542) & (0.525) & (0.525) & (0.523) \\
Out of labor force &  & 0.0943 & 0.164 & -0.0715 & 0.0233 & 0.00933 & -0.00632 \\
 &  & (0.455) & (0.484) & (0.486) & (0.481) & (0.484) & (0.488) \\
Unemployed &  & 0.752 & 0.611 & 0.273 & 0.301 & 0.301 & 0.299 \\
 &  & (0.592) & (0.571) & (0.568) & (0.538) & (0.542) & (0.543) \\
\multicolumn{8}{l}{\textbf{\underline{Location-based controls}}}\\
Population density &  &  & 0.0320 & 0.0409 & 0.0429 & 0.0435 & 0.0430 \\
 &  &  & (0.0334) & (0.0300) & (0.0293) & (0.0294) & (0.0293) \\
Cumulative cases &  &  & 0.00572 & 0.00570 & 0.00444 & 0.00450 & 0.00470 \\
 &  &  & (0.00475) & (0.00442) & (0.00418) & (0.00418) & (0.00423) \\
Daily deaths &  &  & -0.001131 & -0.004514 & -0.003625 & -0.00366 & -0.00382 \\
 &  &  & (3.626) & (3.576) & (3.627) & (3.632) & (0.00371) \\
Stay-at-home order &  &  & -0.355 & -0.343 & -0.446 & -0.451 & -0.433 \\
 &  &  & (0.501) & (0.492) & (0.452) & (0.456) & (0.466) \\
Region controls & No & No & Yes & Yes & Yes & Yes & Yes \\
 &  &  &  &  &  &  &  \\

 \pagebreak
\multicolumn{8}{l}{\textbf{\underline{Preference controls}}}\\
Bomb risk score &  &  &  & -0.0339*** & -0.0296*** & -0.0296*** & -0.0300*** \\
 &  &  &  & (0.00934) & (0.00889) & (0.00896) & (0.00885) \\
Prosocial values &  &  &  & 1.564*** & 1.485*** & 1.495*** & 1.503*** \\
 &  &  &  & (0.338) & (0.334) & (0.335) & (0.334) \\
Protect others &  &  &  & 1.512*** & 0.918*** & 0.926*** & 1.017** \\
 &  &  &  & (0.332) & (0.342) & (0.346) & (0.395) \\
Quiz attempts &  &  &  &  & 0.189 & 0.197 & 0.197 \\
 &  &  &  &  & (0.170) & (0.171) & (0.171) \\
\multicolumn{8}{l}{\textbf{\underline{Ideology}}}\\
Conservative &  &  &  &  & -0.282*** & -0.250*** & -0.207** \\
ideology &  &  &  &  & (0.0565) & (0.0591) & (0.0995) \\
Fine--ideology &  &  &  &  &  & -0.110* & -0.110* \\
interaction &  &  &  &  &  & (0.0618) & (0.0618) \\
Nudge--ideology &  &  &  &  &  & -0.0266 & -0.0265 \\
interaction &  &  &  &  &  & (0.0447) & (0.0448) \\
Residual &  &  &  &  &  &  & -0.0663 \\
 &  &  &  &  &  &  & (0.120) \\
Constant & 1.456*** & -3.547** & -2.753* & -2.927* & -2.051 & -2.213 & -2.415 \\
 & (0.312) & (1.611) & (1.627) & (1.579) & (1.670) & (1.666) & (1.700) \\
 &  &  &  &  &  &  &  \\
Observations & 8,000 & 8,000 & 8,000 & 7,880 & 7,880 & 7,880 & 7,880 \\
No of subjects & 400 & 400 & 400 & 394 & 394 & 394 & 394 \\
\bottomrule
\end{longtable}
\end{ThreePartTable}
}

It is important to bear in mind that subjects were not randomly allocated to their ideology (as they were to the experimental treatments) -- clearly, that would not be feasible. As such, the association we have found between ideology and social distancing decisions may not be causal -- it is possible that ideology is \emph{endogeneous}. That is, it could be correlated with some unobserved factor that affects social distancing decisions.

Therefore, we use an instrumental variables approach to deal with possible endogeneity. We use a measure of subjects' skepticism of global warming as the instrument. As a partisan issue in the United States, this is strongly correlated with political ideology \citeplatex{mccright2011politicization}. However, as it is unrelated to COVID-19 and the types of decisions that we ask subjects to make in this experiment, it should not have a \emph{separate} effect on social distancing decisions. 

Mathematically, this means that after we have controlled for ideology, then skepticism of global warming is uncorrelated with the error term, $\epsilon$, in the regression. This property is required for the instrument to be \emph{valid}, and so to give consistent estimates. Note that it is not possible to test this -- it is an assumption.\footnote{See, for example, \citelatex{wooldridge2010econometric}.}

As we are using a Logit specification, it is not possible to do standard 2-Stage Least Squares. Instead we use a Control Function method \citeplatex{train2009discrete}. The Control Function method takes predicted residuals from the first stage regression, and adds them into the Logit regression. This is in contrast to 2-Stage Least Squares, which takes predicted values from the first stage, and uses them instead of the (potentially) endogenous variable in the second stage.

Model 7 of \cref{tab:logit_main} reports the preferred regression specification with the instrumental variables method. Using the instrument does not have a large impact on the point estimate for the ideology variable. This suggests that there may in fact be a causal relationship between ideology and social distancing decisions -- conservatives practice less social distancing.

\textbf{Partial Effects.} Using our preferred Logit specification -- model 6 of \cref{tab:logit_main} -- we can predict the probability that an agent chooses to practice social distancing. With estimated regression coefficients $\hat{\beta}$ and a set of subject characteristics $x$, then:
\begin{align}
    Pr(y_{it}=1 | x_{it}) = \frac{ e^{ \hat{\beta}^{'} x } }{1 + e^{ \hat{\beta}^{'} x } }.
\end{align}

However, calculating partial effects with a Logit model is far from straightforward. In a non-linear model, the partial effect of one variable depends on the full set of an individual's characteristics, and so is highly heterogeneous. For example, the estimated partial effect of age depends on an agent's gender, race, religion, degree of risk aversion -- and \emph{all other} variables in the model. Note that this even includes variables that are not statistically significant.

Given this, we calculate a variant on Average Partial Effects (APE). This is more useful than a Partial Effect at the Average; especially given our extensive use of binary variables -- a Partial Effect at the Average would give us the partial effect for a subject who is 47\% female, 85\% white, and at a node position that simply does not exist (50\% `close-knit', 40\% peripheral, and 10\% superspreader). 

To calculate our variant on an APE for a variable, for example, gender, we predicted the probability of social distancing for each of our 400 subjects, first assuming that they are all male, and then second assuming that they are all female. This gives us an individual partial effect of gender for each subject -- taking an average across all subjects yields the variant on an APE. We assume that subjects are not exposed to a policy intervention (fine/nudge) when calculating APEs for all (other) variables. When the variable is continuous, we first assume that all subjects are at the 25th percentile of that variable (based on the actual distribution in the experiment), and second that they are at the 75th percentile.

We use this variant -- only looking at the 400 subjects in our experiment -- due to data constraints. While data is readily available on the population-wide distributions of most of the individual variables, they are only available as marginal distributions, not as joint distributions. That is, one can easily find an age distribution, a gender distribution, and an education distribution for the US population; but they are only available separately. A joint distribution -- especially over \emph{all} of the variables in our preferred specification is not available. Therefore, our variant allows us to calculate partial effects for individuals who actually exist, and so makes our APE meaningful. Figure \ref{fig:3} in the main text shows a box plot Partial Effects for variables that are statistically significant.

\subsubsection{Convergence}\label{sec:robustness_convergence}

The analysis above focuses on the last 10 rounds of Baseline and Intervention. Here, we validate this choice by showing that the majority of participants exhibit clear convergence behavior after the first 10 rounds in both parts of the experiment. 
We define individual convergence as follows:

\begin{definition}
    A participant converges to a strategy $s$ by round $n$ if (i) she used this strategy for the last $k$ rounds (including $n$), and (ii) in all subsequent rounds $[n+1, 20]$ the number of consecutive deviations from the chosen strategy does not exceed $a$.
\end{definition}

We consider three types of convergence strategies as follows. In both networks, we look at the strategy where the subject always chooses the same action. For the star network, we also consider two extra strategies. In one strategy the participant always chooses the same action when she is in the superspreader position and the complement action when she is peripheral. In the other strategy, she always chooses the same action when she is the superspreader and alternates between the actions when peripheral.

We set $k = 4$ and $a=2$. This means that the earliest a participant can be considered to converge to a particular strategy is by round 4 if she has not deviated from this strategy in the last four rounds, and in all subsequent rounds, she never performs more than two consecutive deviations.

Using the above definition, by round 11 at least 80\% of our participants have converged to a certain strategy in all parameterizations.

\begin{figure}[t]
    \centering
    \includegraphics[width=\textwidth]{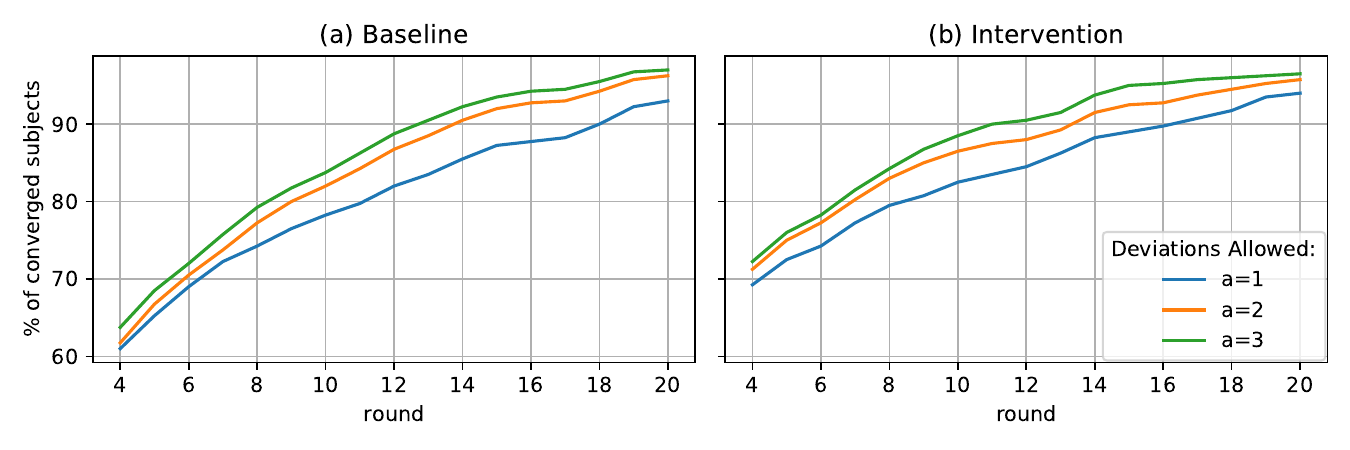}
    \caption{Evolution of the share of converged participants throughout the xxperiment separately for Baseline and Intervention, $a \in [1,3]$. Note: we exclude the first 3 rounds in both parts since $k=4$.}
    \label{fig:convergence}
\end{figure}

As a robustness check, we also consider $a=1$ and $a=3$, allowing for one and three consecutive deviations respectively. \Cref{fig:convergence} plots the share of converged participants for each round separately by parts for $a \in [1,3]$. We can see that the share of converged subjects does not change much when we allow for a more/less conservative definition. In particular, with $a=1$ the share of subjects who converge by round 11 in Baseline drops to 79.8\% while in Intervention it reaches 83.5\%. With $a=3$ the share of subjects who converge by round 11 in Baseline and Intervention stands at 86.3\% and 90\% respectively.

The above analysis suggests that it is reasonable to claim that the absolute majority of subjects converge to a particular strategy by round 11 in both parts of the experiment.

\subsubsection{Perception and Behavior} \label{sec:analysis_perception}

As part of the post-experimental questionnaire, subjects are asked whether they agree that fines and nudges are effective in promoting social distancing. These two questions are both on a 5-point Likert scale, with higher values indicating greater disagreement with the statements. \Cref{tab:attitudes} summarizes individual self-reported perceptions of fines and nudges in our sample. The majority of subjects believe that nudges are effective in promoting social distancing, while attitudes to fines appear to be polarized.

\begin{table}[ht]
\centering
\caption{Self-reported perception of fines and nudges, $n = 400$.}
\label{tab:attitudes}
\begin{tabular}{@{}lccccc@{}}
\toprule
 & strongly agree & agree & \begin{tabular}[c]{@{}c@{}}neither agree\\ nor disagree\end{tabular} & disagree & strongly disagree \\ \midrule
fines are effective & 11.75\% & 35.5\% & 21.25\% & 20.25\% & 9.25\% \\
nudges are effective & 31\% & 50.5\% & 13.75\% & 3.5\% & 1.25\% \\ \bottomrule
\end{tabular}
\end{table}

We construct a group-level index for the perception of fines, equal to one if at least three subjects in the group believe that fines are effective, i.e. they (strongly) agree with the corresponding statement, and an identical index for the perception of nudges. We then construct another index that captures changes in the observed behavior. This index is equal to one if the average level of social distancing in the last 10 rounds of Intervention is higher than in the last 10 rounds of Baseline. We then compare the perception indices to the behavior index.

\begin{figure}[ht]
    \centering
    \includegraphics[width=0.5\textwidth]{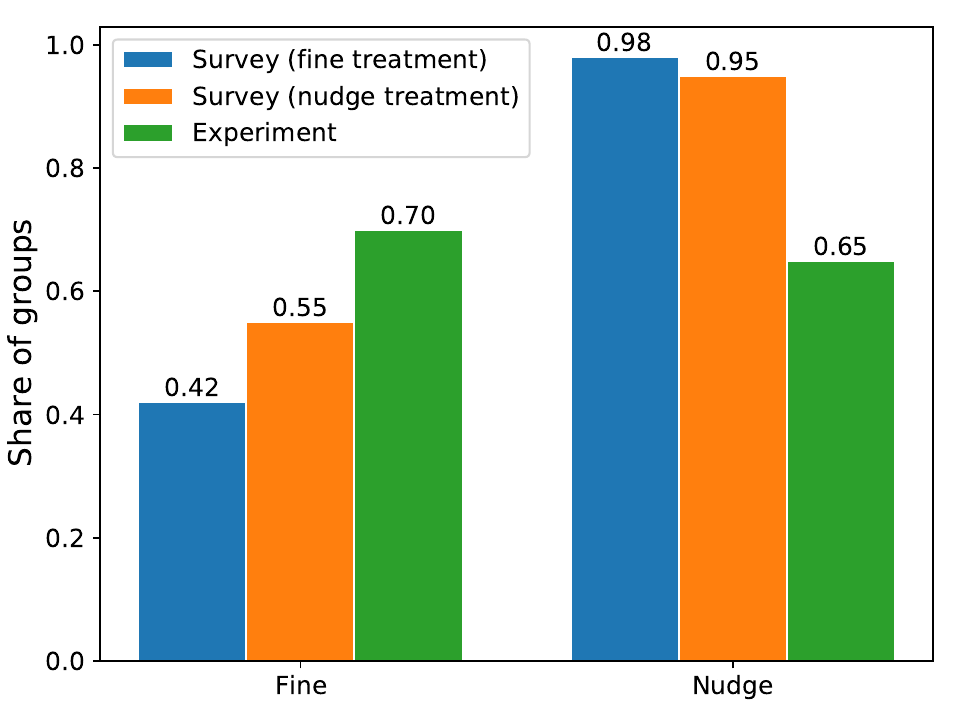}
    \caption{Perception of effectiveness of fines and nudges versus observed effectiveness.}
    \label{fig:perception_behavior}
\end{figure}

\Cref{fig:perception_behavior} summarizes how group-level perception of fines and nudges compares to behavior. Two patterns emerge. First, subjects seem to underestimate the effectiveness of fines: only 42\% of groups in the fine treatment (55\% in the nudge treatment) believe in the effectiveness of fines, whereas 70\% of groups subjected to fines show an increase in average distancing levels. Second, participants seem to overestimate the effectiveness of nudges -- 95\% of the groups in the nudge treatment (98\% in the fine treatment) believe that nudges are effective, while only 65\% of the groups actually increase their average distancing levels when subjected to nudges. 

To test this formally, we use non-parametric analysis with the three group-level indices. We find that irrespective of treatment they were placed in, subjects' perception of fines matches their (fines') measured effectiveness (WSR and MW, $p=0.99$ and $p > 0.92$ respectively). On the other hand, participants overestimate the effectiveness of nudges (WSR and MW, $p=0.001$ $p<0.0001$ respectively).

\bibliographystylelatex{chicago}  
\bibliographylatex{appendix_bib.bib} 

\end{document}